\documentclass[12pt,preprint]{aastex}

\shorttitle{GALAXY SIZE EVOLUTION}
\shortauthors{I. TRUJILLO ET AL}

\begin{document}

\title{THE LUMINOSITY--SIZE AND MASS--SIZE RELATIONS OF GALAXIES OUT TO z $\sim$
3\altaffilmark{1}}
\author{IGNACIO TRUJILLO\altaffilmark{2}, GREGORY RUDNICK\altaffilmark{3}, HANS--WALTER RIX\altaffilmark{2},
 IVO LABB\'E\altaffilmark{4}, MARIJN
FRANX\altaffilmark{4}, EMANUELE DADDI\altaffilmark{5}, PIETER G. VAN DOKKUM\altaffilmark{6}, NATASCHA M. F\"ORSTER
SCHREIBER\altaffilmark{4},
KONRAD KUIJKEN\altaffilmark{4}, ALAN MOORWOOD\altaffilmark{5}, HUUB R\"OTTGERING\altaffilmark{4},
 ARJEN VAN DE WEL\altaffilmark{4}, PAUL VAN DER
WERF\altaffilmark{4}, LOTTIE VAN STARKENBURG\altaffilmark{4}}

\altaffiltext{1}{Based on observations collected at the European Southern Observatory,
Paranal, Chile (ESO LP 164.O--0612). Also, based on observations with the
NASA/ESA $Hubble$ $Space$ $Telescope$, obtained at the Space Telescope Science
Institute, which is operated by AURA Inc, under NASA contract NAS 5--26555.}
\altaffiltext{2}{Max-Planck-Institut f\"ur Astronomie, K\"onigstuhl 17, 69117
Heidelberg, Germany}
\altaffiltext{3}{Max-Planck-Institut f\"ur
Astrophysik, Postfach 1317, D-85748, Garching, Germany}
\altaffiltext{4}{Leiden Observatory, P.O. Box 9513, NL--2300 RA, Leiden,
 The Netherlands}
\altaffiltext{5}{European Southern Observatory, D--85748,
Garching, Germany}
\altaffiltext{6}{Department of Astronomy, Yale University, P.O. Box 208101,
New Haven, CT 06520-8101, USA}

\begin{abstract}

The luminosity--size and stellar mass--size distributions of galaxies out to
z$\sim$3 is presented. We use very deep near--infrared images of the Hubble
Deep Field South in the J$_{s}$, H, and K$_{s}$ bands, taken as part FIRES at
the VLT, to follow the evolution of the optical rest--frame sizes of galaxies.
For a total of 168 galaxies with K$_{s,AB} \leq$23.5, we find that the
rest--frame V--band sizes r$_{e,V}$ of luminous galaxies
($<$L$_V$$>$$\sim$2$\times$10$^{10}$h$^{-2}$L$_\odot$) at 2$<$z$<$3 are 3 times
smaller than for equally luminous galaxies today. In contrast, the mass--size
relation has evolved relatively little: the size at mass
$<$M$_\star$$>$$\sim$2$\times$10$^{10}$h$^{-2}$M$_\odot$, has changed by
20($\pm$20)\%  since z$\sim$2.5. Both results can be reconciled by the fact
that the stellar M/L ratio is lower in the luminous high z galaxies than in
nearby ones because they have young stellar populations. The lower incidence of
large galaxies at z$\sim$3 seems to reflect the rarity of galaxies with high
stellar mass.

\end{abstract}


\section{INTRODUCTION}

 The size evolution of galaxies with redshift serves as an important constraint
 on models of galaxy evolution.  In the current standard cosmology
 ($\Omega_M$=0.3, $\Omega_{\Lambda}$=0.7), hierarchical models of galaxy
 formation predict a strong increase in the characteristic size of  galaxies
 since z$\sim$3 (Baugh et al. 1998; Mao, Mo \& White 1998; Avila--Reese \&
 Firmani 2001; Somerville, Primack \& Faber 2001).  This, however, has not yet
 been extensively tested by observations.  Early studies using ground--based
 telescopes (Smail et al. 1995) and the Hubble Space Telescope (HST; Casertano
 et al. 1995) showed that at magnitudes of $I\approx 22$ and $R\approx 26$,
 where the expected median redshift is greater than 0.5, the dominant field
 population is formed by very small systems with a mean scale length of
 $\sim0\farcs2-0\farcs3$. These objects are more compact than one would expect
 by assuming a fixed intrinsic physical size (Smail et al.  1995). Subsequent
 studies (Roche et al. 1998) at fainter magnitudes ($22<I<26$) suggested that
 most size evolution occurs at $z>1.5$.

 The study of galaxy properties  at larger redshifts ($z>2$) was dramatically
 improved by the identification of a large population of star--forming galaxies
 (Steidel et al. 1996). These Lyman--Break galaxies (LBGs) are identified by
 the redshifted break in the far UV continuum caused by intervening and
 intrinsic neutral hydrogen absorption.  Sizes of galaxies at $z\sim$3 have
 been measured for LBGs (Giavalisco, Steidel \& Maccheto 1996; Lowenthal et
 al.  1997; Ferguson et al. 2003), but in the rest--frame UV part of their
 spectra. In the UV these galaxies appear compact ($r\sim0\farcs2-0\farcs3$,
 $\sim$1.5--3 h$^{-1}$ kpc), in good qualitative agreement with the predictions
 for the build--up of  stellar mass from hierarchical formation scenarios (Mo,
 Mao \& White 1999).  However, the selection technique and the observed
 rest--frame wavelength raises the following question: are the galaxies
 selected (Lyman--break galaxies) and the sizes measured (UV sizes)
 representative of the radial stellar mass distribution of the luminous high--z
 galaxy population? Put differently, is the radial extent of the instantaneous,
 relatively unobscured star formation, which is measured by the rest--frame far
 UV light, indicative of the radial extent of the stellar mass distribution?

 To properly test the model predictions one would ideally like to trace the
 size evolution of galaxies in the optical (rather than UV) rest--frame at
 every redshift.  Any observed size evolution would then reflect true
 evolutionary changes  not subject to  the changing appearance of galaxies in
 different bandpasses, an effect known as the morphological $k$-correction. 
 Most of the past studies using constant rest--frame bands have been limited to
 modest redshifts (z$\lesssim$1; e.g. Lilly et al. 1998) due to the dearth of
 very deep near infrared images which allow one to reach the rest--frame
 optical.

 To map the size evolution of the stellar body of galaxies it is necessary to
 conduct an analysis at wavelengths at least as red as the rest--frame
 optical.  At $z\gtrsim0.8$ this implies selecting and analyzing galaxies from
 very deep near--infrared images. In this paper we use data for the Hubble Deep
 Field South from the Faint InfraRed Extragalactic Survey (FIRES; Franx et al. 
 2000) to address this issue\footnote{The size properties of galaxies in the
 MS1054--03 FIRES field (F\"orster Schreiber et al. in preparation) will be
 discussed in a forthcoming paper.}. We use these data to constrain the size
 evolution, i.e. we test whether for a given rest--frame luminosity, or a given
 stellar mass,  the sizes of the high--z population are equal to or different
 from those of nearby galaxies?  To assess the degree of evolution, if any, it
 is crucially important that good local calibrating data be available.  With
 the advent of large local surveys (e.g. the Sloan Digital Sky Survey (SDSS);
 York et al. 2000), we now have complete samples of local galaxies with
 accurate measurements of fundamental properties such as luminosity or size to
 use as low redshift reference points.

 This paper is organized as follows: the data and the measurement
 technique are described in \S2; in \S3 we present the
 luminosity--size and mass--size relation of the high--z galaxies and
 discuss how selection effects play a role in interpreting the
 observed trends; \S4 describes simulations of how the local galaxy
 population (as provided by the SDSS data) would appear at high--z. By
 comparing with the FIRES data, we constrain the size evolution for
 the galaxies in our sample.  Finally, in Sect. 5 we discuss our
 results.

\section{OBSERVATIONS, DATA, AND SIZE DETERMINATIONS}

\subsection{Data and catalog construction}

\begin{figure}
\plotone{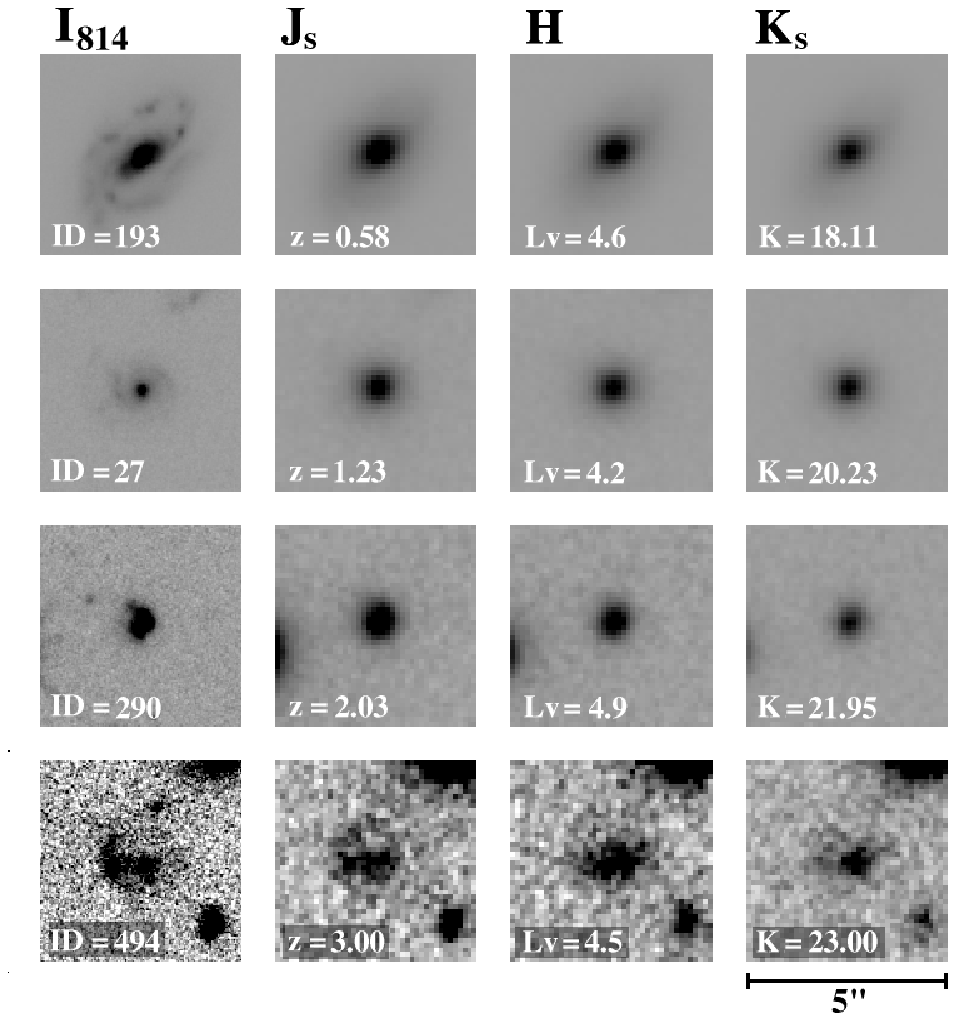}
 \caption{A mosaic of similar rest--frame V--band luminosities
 galaxies at different redshifts. The apparent K-band magnitude
 decreases towards the bottom.  The luminosities are given in
 10$^{10}$ solar luminosities. The galaxies are shown in four
 different filters.}
\label{mosaic}
\end{figure}

 Ultradeep near--infrared images of the Hubble Deep Field South were obtained
 as part of the FIRES survey and the data processing and photometry are
 discussed in detail by Labb\'e et al. (2003)\footnote{The catalog and reduced
 images are available at http://www.strw.leidenuniv.nl/$\sim$fires}.  Briefly,
 using ISAAC (Moorwood 1997; field of view of 2.5$'\times$2.5$'$ and pixel
 scale $0\farcs119$)\footnote{The ISAAC pixel scale is actually $0\farcs147$,
 however we resampled the ISAAC pixels to 3x3 blocked HDF-S WFPC2 pixels.} on
 the VLT the HDF--S was imaged for 33.6 hours in J$_s$, 32.3 hours in H, and
 35.6 hours in K$_s$. The effective seeing in the reduced images is
 approximately $0\farcs47$ in all bands. The depth (3$\sigma$) reached was 26.8
 mag in J$_s$, 26.2 mag in H, and 26.2 mag in K$_s$ for point sources. All
 magnitudes in this paper are given in the AB system unless stated explicitly
 otherwise. Some examples of galaxies in these ultra--deep images are presented
 in Fig. \ref{mosaic}.  Combining these near infrared data with deep optical
 HST/WFPC2 imaging (version 2; Casertano et al. 2000), we assembled a
 K$_s$--selected catalogue containing 833 sources, of which 624 have
 seven--band photometry, covering 0.3--2.2$\mu$m.  Stars were identified, and
 removed from the catalog, if their spectral energy distributions (SEDs) were
 better fitted by a single stellar template than by a linear combination of
 galaxy templates.  Four of the stellar candidates from this color
 classification were obviously extended and were reidentified as galaxies.  Two
 bright stars were not identified by their colors because they are saturated in
 the HST images and were added to the list by hand. Photometric redshifts were
 estimated for the catalogued galaxies following Rudnick et al. (2001) (see
 also  Section 3).

 The sample of galaxies is selected in the K$_s$--band. For z$\lesssim$3 this
 filter reflects galaxy flux at wavelengths redward than the rest--frame
 V--band and so selects galaxies in a way that is less sensitive to their
 current unobscured star formation rate than selection in the rest-frame UV. To
 select galaxies with reliable photometry, we exclude the much less exposed
 borders of our combined K$_s$ ``dithered'' image (see Labb\'e et al. 2003),
 taking only those galaxies whose fractional exposure time is $\ge$35\% of the
 maximum.  To ensure sufficient signal-to-noise for the subsequent size
 determinations, we limit ourselves to the 171 objects with K$_s\leq23.5$ and
 with ISAAC and WFPC2 coverage.


\subsection{Measuring sizes}
\label{sizemeas}

The  multiband imaging allows us to make a homogeneous comparison of the
 rest--frame optical size for all sample galaxies at redshift z$\lesssim$3. We
 measure the sizes of galaxies at all redshifts consistently by fitting the
 profile of each galaxy in the band--pass which corresponds most closely to the
 rest--frame V--band at that redshift: for 0$<$z$<$0.8 we fit the
 I$_{814}$--band, for 0.8$<$z$<$1.5 the J$_s$--band, for 1.5$<$z$<$2.6 the
 H--band, and for 2.6$<$z$<$3.2 the K$_s$--band.

 At high redshift the angular sizes of typical galaxies in our sample are
 comparable to the size of the seeing ($\sim0\farcs47$). Consequently, the
 intrinsic structure and size of the galaxies must be obtained by adopting a
 surface brightness (SB) model and convolving it with the image PSF. This
 approach is well tested and successful at fitting low--z galaxies.

 We seek a flexible parametric description of the galaxies' SB distribution,
 without resorting to multi--component models. The population of galaxies at
 any redshift is likely a mixture of spirals, ellipticals and irregular
 objects. Elliptical galaxies (from dwarfs to cDs) are well fitted by a
 S\'ersic model $r^{1/n}$ (S\'ersic 1968), as demonstrated by a number of
 authors (see e.g. Trujillo, Graham \& Caon 2001 and references therein). The
 S\'ersic model is given by: \begin{equation}
 I(r)=I(0)\exp^{-b_n(r/r_e)^{1/n}}, \end{equation} where $I(0)$ is the central
 intensity and $r_e$ the effective radius, enclosing half of the flux from the
 model light profile.  The quantity $b_n$ is a function of the radial shape
 parameter $n$ -- which defines the global curvature in the luminosity profile
 -- and is obtained by solving the expression $\Gamma (2n)$$=$$2\gamma
 (2n,b_n)$, where $\Gamma(a)$ and $\gamma(a,x)$ are respectively the gamma
 function and the incomplete gamma function.

 The disks of spiral galaxies are also well described by a S\'ersic model with
 n=1, corresponding to an exponential profile.  The S\'ersic model (with its
 free shape parameter $n$)  is flexible enough to fit the radial profiles of
 nearly every galaxy type\footnote{Even early--type spirals composed by a bulge
 plus a disk can be fitted by a single S\'ersic model (Saglia et al. 1997).}.
 For this reason and for its simplicity, we decided to use it for measuring the
 sizes of galaxies in our data.

 Both the intrinsic ellipticities of the galaxies and the effects of the seeing
 on the images were taken into account when fitting the model. Details of the
 particular model fitting method are given in Trujillo et al. (2001) and
 Aguerri \& Trujillo (2002).

\begin{figure}
\plotone{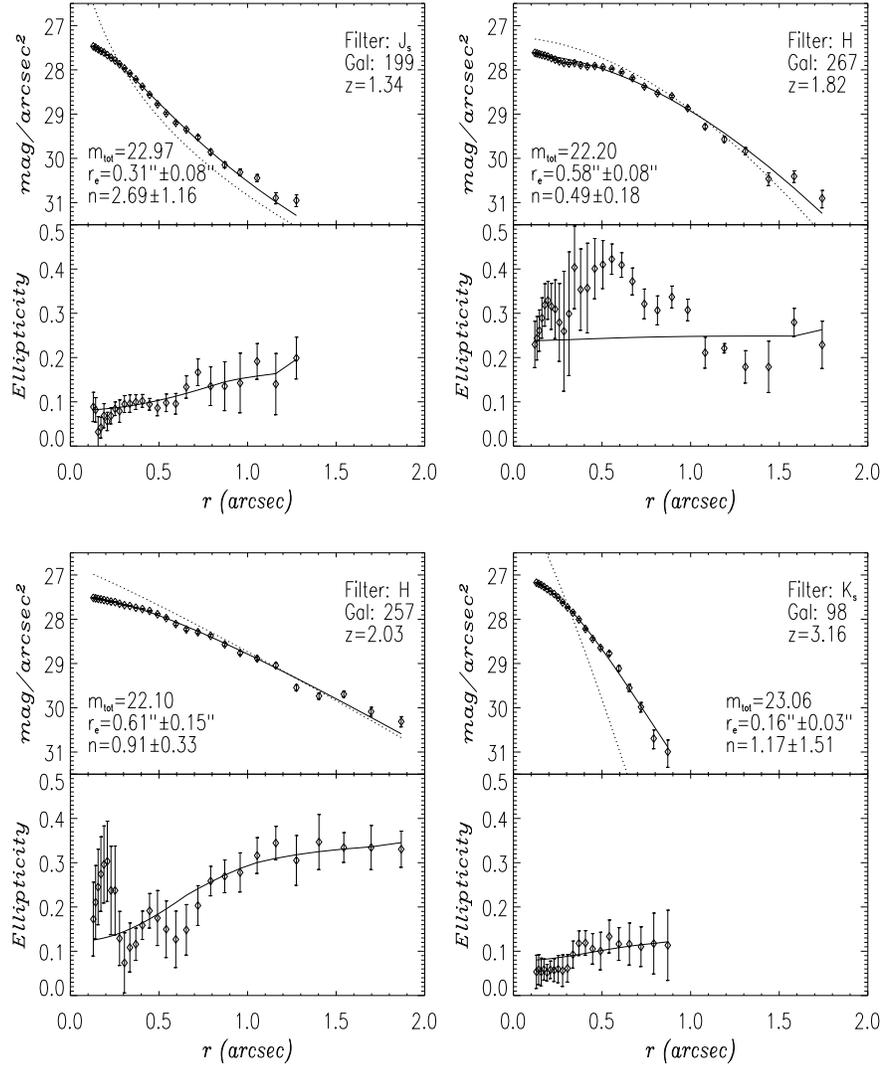}
\vspace{-4cm}
 \caption{\scriptsize Surface brightness
 and ellipticity semi--major radial profiles fitting to four galaxies in our sample. Every
galaxy is fitted in its optical rest--frame band following the criteria
explained in the text.  The galaxy identification
numbers correspond to the catalog identification  (see Labb\'e et al.
(2003)). Superimposed on the surface brightness profile data are the model
profile (dashed line) and the convolution of this model profile (solid line) to
match the data. The solid lines in the ellipticity radial profiles show the fit
to the ellipticities using our algorithm. Intrinsic ellipticities (i.e. not
seeing convolved) of the galaxies can be obtained by extrapolating to infinity
the solid lines. Details of the fitting algorithm can be found in
Trujillo et al. (2001) and Aguerri \& Trujillo (2002).}
\label{ajustes}
\end{figure}

\begin{figure}

\plotone{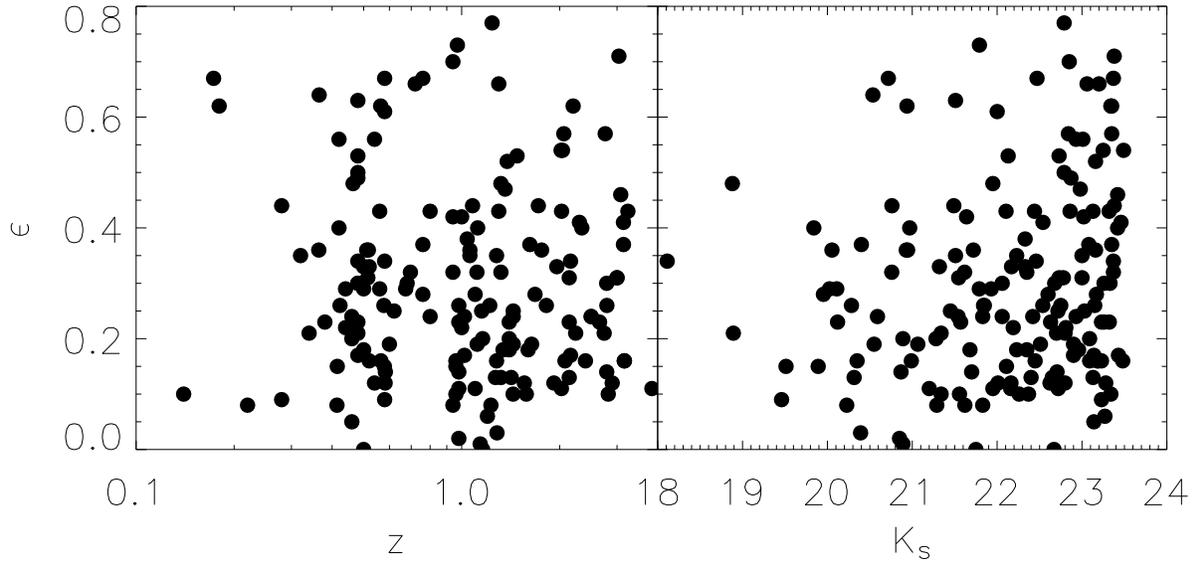}

\caption{The intrinsic  (i.e. the recovered non--seeing affected) ellipticity
of the galaxies versus: a) the redshift of the observed sources and b) the
apparent K$_s$ total magnitudes. No clear relation is observed.}

\label{ellipticity}

\end{figure}

 We start by measuring the SB and ellipticity profiles  along the  major radial
axis by fitting isophotal ellipses to the sample object images, using the  
task ELLIPSE within IRAF\footnote{IRAF is distributed by the National Optical
Astronomical Observatories, which are operated by AURA, Inc. under contract to
the NSF.}. The fits extend down to 1.5 times the standard deviation of the sky
background of the images. Some examples of the model fits to our sample
galaxies are presented in Fig.~\ref{ajustes}. A Levenberg--Marquardt
non--linear fitting algorithm (Press et al. 1992) was used to determine the set
of free parameters which minimizes $\chi^2$. To do this, we  fit simultaneously
the observed 1D  and ellipticity  profiles using a PSF convolved model for
each. In what follows,   we  refer to the \textit{circularized effective
radius} of the fitted model, i.e. $r_e=a_e\sqrt{(1-\epsilon)}$, as the
\textit{size} of the galaxies; here $a_e$ is the semi--major effective radius 
and $\epsilon$ the projected ellipticity of the galaxy model. We checked that
the estimate of the intrinsic ellipticity of our sources, and hence the
conversion to circularized effective radius, was not systematically affected by
the seeing,  by searching for trends with z or the K$_s$ apparent magnitude
(Fig. \ref{ellipticity}).  No significant trends were found.

\begin{figure}
\plotone{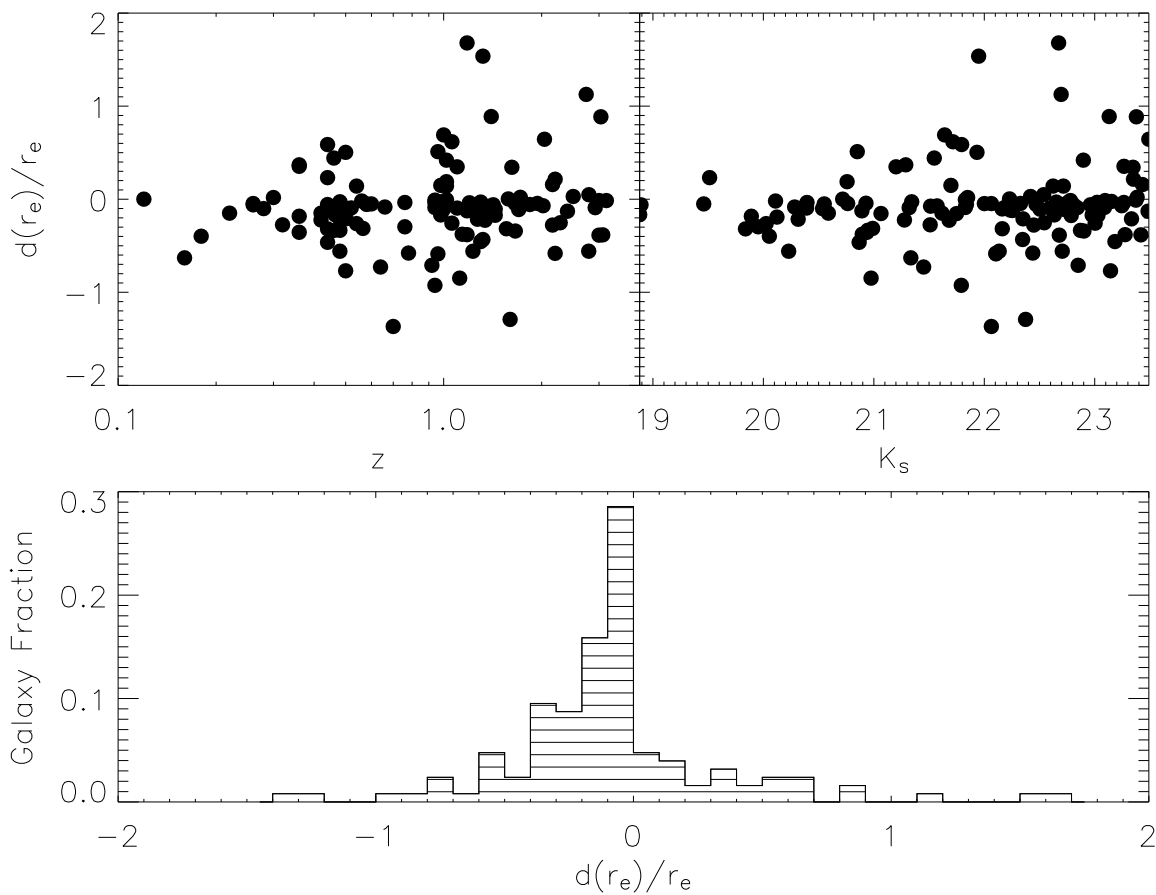}

\caption{The relative error between the size estimation using our code and
GALFIT
d(r$_e$)/r$_e$=2$\times$(r$_e$(GALFIT)-r$_e$(ours))/(r$_e$(GALFIT)+r$_e$(ours))
is shown versus $z$ and versus the apparent K$_s$ magnitude. No clear
trend is found. The histogram shows that for $\sim$68\% of the
galaxies the difference is less than 35\%.}

\label{recomparison}

\end{figure}

  The PSF was estimated for every band by fitting a Moffat function to star
 profiles. We find the following $\beta$ and FWHM (Full at Width Half Maximum)
 as our best fitting stellar parameters: $\beta$=2.5, FWHM=$0\farcs147$
 (I$_{814}$--band); $\beta$=3, FWHM=$0\farcs46$ (J$_s$--band); $\beta$=3,
 FWHM=$0\farcs49$ (H--band), and $\beta$=3, FWHM=$0\farcs47$ (K$_s$--band). 
 When fitting objects close to the resolution limit, it is crucial to have an
 accurate measure of the PSF. To test the robustness of our size measurements
 against slight errors in our PSF determination we compared our sizes to those
 determined using a completely independent fitting algorithm (GALFIT; Peng et
 al. 2002).  This code uses the 2D profiles of the stars themselves to convolve
 the models with the seeing.  In Fig.~\ref{recomparison} we show the relative
 error between the size estimates using both our code and GALFIT.  The
 difference between the sizes does not show any clear trend with $z$ or the
 apparent K$_s$ magnitude. The agreement between these two different algorithms
 ($\sim$68\% of the galaxies have a size difference less than 35\%)
 corroborates the robustness of the size determination.  One reason for this is
 that our very deep NIR data allows us to sample the profiles out to
 approximately 2 effective radii, providing ample constraints for the fit.  For
 the smallest objects, our bright total magnitude limit (K$_s$=23.5) has the
 effect that the measured SB profile extends over 5 magnitudes and is therefore
 very well characterized. The size errors of each galaxy are taken into account
 in the subsequent analysis.

  There are 3 galaxies where the size estimation is ill defined because they
have a close companion. These galaxies represent only $\sim1.5\%$ of the total
and are all at z$<$1.15. The final sample is composed of 168 galaxies. The sizes
of these galaxies are shown in Table \ref{tabdata}.

\begin{figure}
\plotone{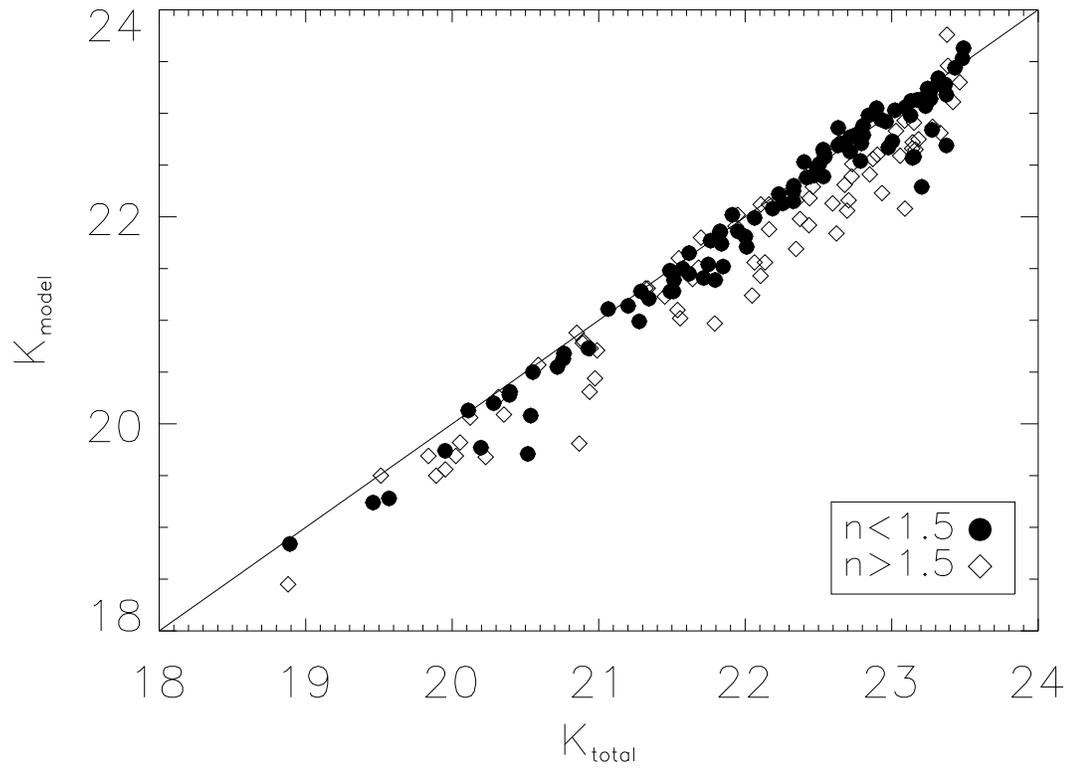}
\caption{The total magnitude retrieved from the model fitting is compared to the
total magnitude measured in a model independent way for the galaxies analyzed in
this paper. Galaxies with n$<$1.5 are represented with solid circles and
galaxies with n$>$1.5 with open diamonds.}

\label{ktotkmod}

\end{figure}

 One way to test the quality  of our model fits is to compare the  model
 magnitude with an aperture isophotal magnitude  (Labb\'e et al. 2003) (see
 Fig.  \ref{ktotkmod}). The total luminosity, $L_T$, associated with an
 $r^{1/n}$ profile extended to infinity can be written as \begin{equation}
 L_T=I(0)r_{e}^2\frac{2\pi n}{b_n^{2n}}\Gamma(2n) \label{eqlum1} \end{equation}
 As expected for an extrapolation to infinity, the total model magnitude is
 almost always equal or brighter than the model--independent determination.   
 In general, there is relatively good agreement between the two measures: a
 magnitude difference $<$0.2 mag for 75\% of the sample. The difference between
 the two estimators is largest for objects with the highest n values, as
 expected because of the large amounts of light at large radii in these models
 (Trujillo, Graham \& Caon, 2001).  Galaxies, however, certainly do not extend
 to infinity and the model extrapolation is likely unphysical, especially for
 high-n values.  For this reason we choose the  total luminosity  from Labb\'e
 et al. (2003) in the following analysis.

\section{THE OBSERVED LUMINOSITY/MASS $VS$ SIZE RELATIONS AT HIGH-Z}

\begin{figure}
\plotone{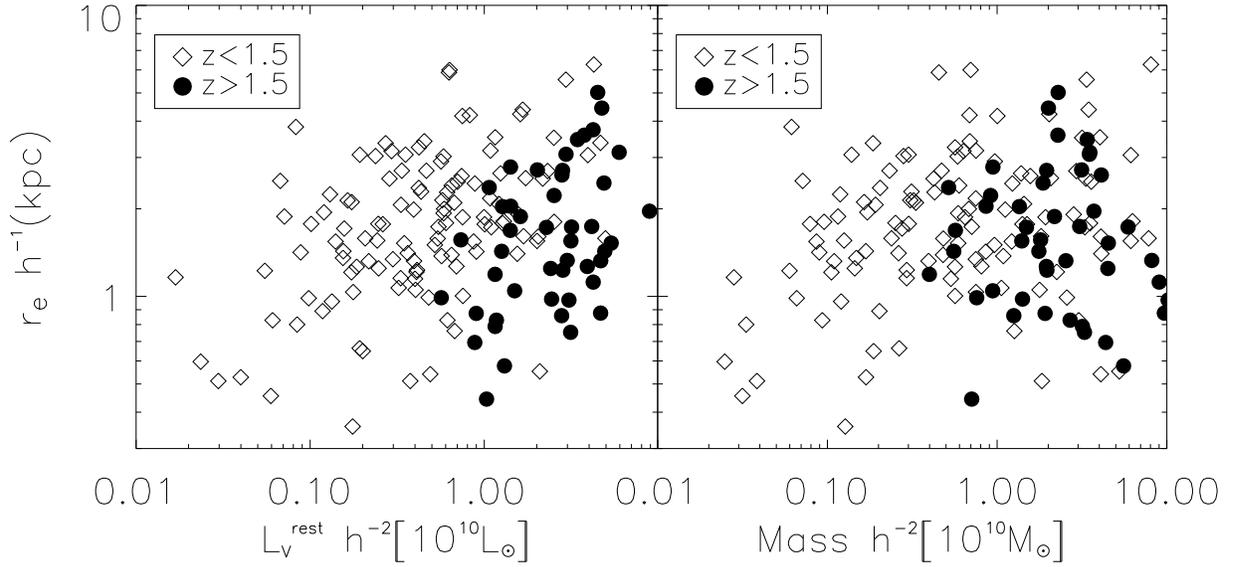}

\caption{$Left$: The distribution of rest--frame optical sizes  versus the
rest--frame V--band luminosities is shown. Galaxies with redshifts smaller than
1.5 are represented with open diamonds and galaxies with redshifts bigger than
1.5 with solid circles. $Right$: Same as before but with the stellar mass. For
clarity error bars are not shown. The mean size relative error is 35\%.}

\label{rawdata}

\end{figure}

 We now present the relations between stellar luminosity, or stellar mass, and
the rest--frame V--band  size over a wide range of redshift.  Throughout, we
will assume  $\Omega_M$=0.3, $\Omega_{\Lambda}$=0.7 and $H_0$=100h km s$^{-1}$
Mpc$^{-1}$.  We convert our measured angular sizes to physical sizes using the
photometric redshift determined for each object\footnote{For 25\% of the
galaxies in our sample $z$ was determined spectroscopically.  When a $z_{spec}$
determination is available, this is the value used.}.  These redshifts, and the
rest-frame optical luminosities, were estimated by fitting a linear combination
of nearby galaxy SEDs and model spectra to the observed flux points (Rudnick et
al. 2001, 2003a).  The accuracy derived from 39 available spectroscopic
redshifts is very good, with $|z_{phot}-z_{spec}| / ( 1 + z_{spec}) \approx
0.05$ for $z>1.4$. A plot of the z$_{spec}$ versus z$_{phot}$ for the present
sample is shown in Labb\'e et al. (2003; Figure 6). We neglect the photometric
redshift uncertainties in our analysis since a redshift error of even $\pm 0.5$
at $z=1.5$ correspond to size errors of $\lesssim 5\%$.  On the other hand, our
photometric redshift uncertainties equate to   $\lesssim$35\% luminosity
errors. For a first analysis step, we have split our sample into a z$\leq$1.5
and a z$>$1.5 bin and plot the rest--frame optical effective radius (the size
estimated in the rest--frame band filter) versus the total luminosity in the
rest--frame V--band in Fig.~\ref{rawdata}.

 We also have explored the relation between stellar mass and size for our
 sample (see Fig. \ref{rawdata}). The stellar mass-to-light ratios (M/L) and
 hence stellar masses for the individual galaxies were estimated by Rudnick et
 al. (2003b) from their rest--frame colors and SEDs, using a technique similar
 to that of Bell \& de Jong (2001).  Briefly, this approach exploits the
 relation between color and stellar mass-to-light ratio (M/L), which exists
 over a wide range of monotonic star formation histories, and which is rather
 robust against the effects of age, dust extinction, or metallicity.  Errors in
 the derived masses will occur in the presence of bursts.  In practice, we
 derive the M/L from the rest-frame ($B-V$) color using the  models presented
 in Bell \& de Jong (2001). We take into account the photometric redshift
 probability distribution and the scatter in the ($B-V$) -- M/L relation when
 calculating our uncertainties (Rudnick et al. 2003b).

\subsection{Selection Effects}

 For studying  galaxy size evolution from Fig. \ref{rawdata}, we must
 understand  the selection effects at play in our sample.  Redshift--dependent
 observational biases can mimic real evolutionary changes in the galaxy
 population, both through biases in the selection of galaxies and through the
 measurement of their sizes.  Knowing the selection effects is also crucial in
 creating mock high redshift catalogs from low redshift surveys.

\begin{figure}
\plotone{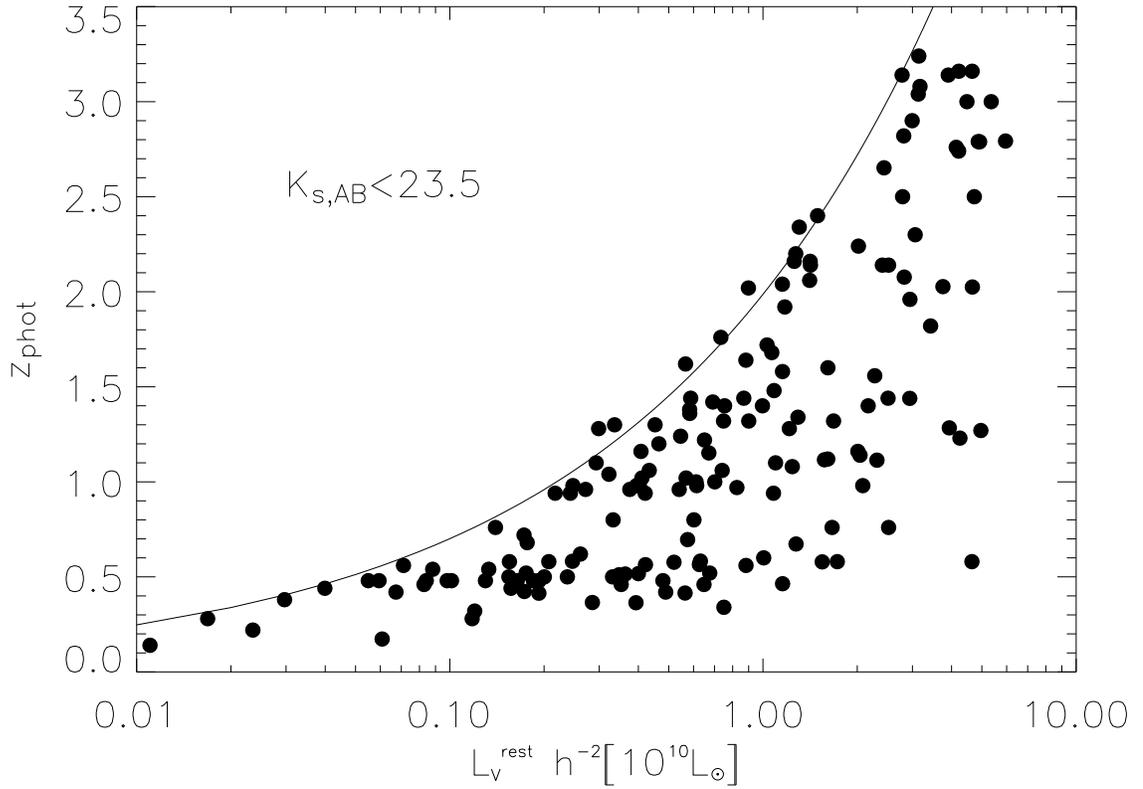}

\caption{The L$_V$--z diagram for the objects selected in our sample with
K$_s<$23.5. The track represent the values of L$_V^{rest}$ for a Scd template
spectra (see Rudnick et al. (2001); Figure 11c) normalized at each redshift to
K$_{s,AB}$=23.5.} 

\label{zphotlum}

\end{figure}

 For a given flux limit (in our case K$_s$=23.5) there is a corresponding
 threshold in the rest-frame luminosity, which increases with redshift. This is
 well illustrated in the L$_V$--z diagram (Fig. \ref{zphotlum}) and
 demonstrates that our high--z sample represents only the most luminous 
 fraction of the galaxy population. The absence of bright galaxies at low
 redshift is largely due to the small volume of the HDF--S over this redshift
 range.

\begin{figure}
\plotone{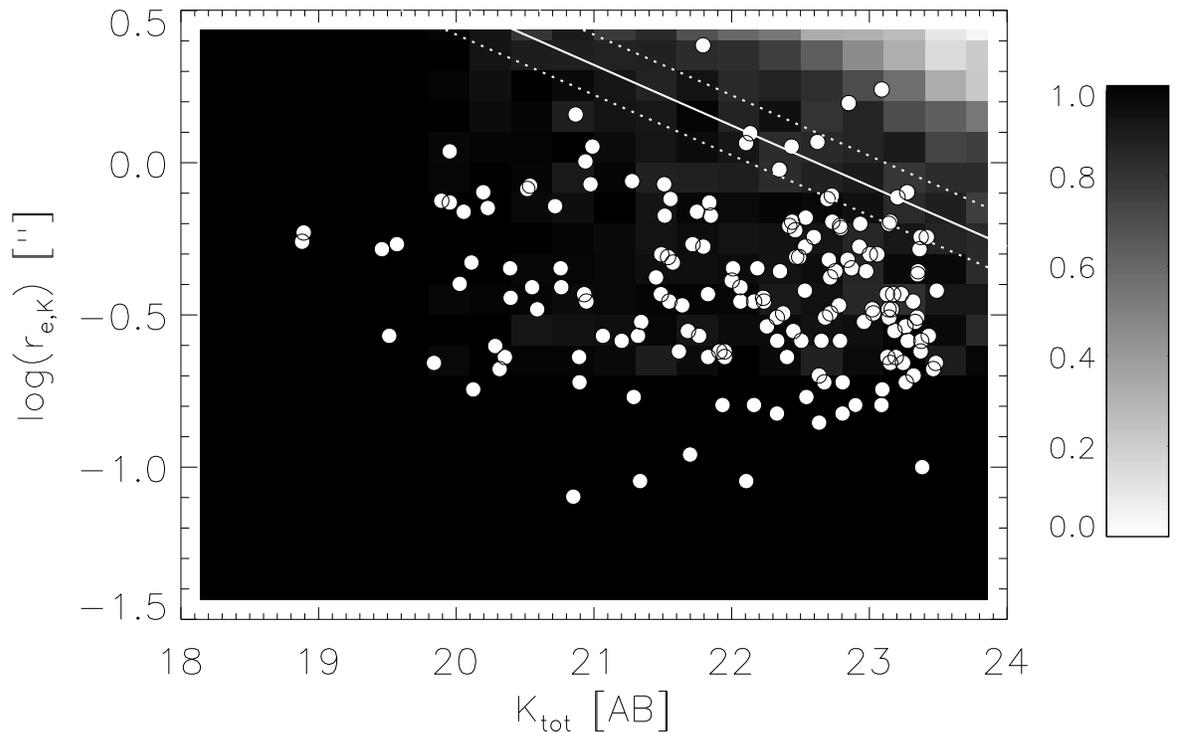}

\caption{A completeness map for detecting galaxies with exponential profiles in
the FIRES K--band data. The horizontal and vertical axis represent the input
values. Overplotted is the K$_s$--band size  versus the apparent K$_s$
magnitude  for the objects in our sample (open points). We  have also shown  
exponential models (n=1) with central surface brightness of 23 and 24
mag/arcsec$^2$ (dotted lines). The solid line represents  the central surface
brightness ($\mu_K(0)$=23.5 mag/arcsec$^2$) at which we are 90\%  complete.} 
\label{ivofig}

\end{figure}

 The detectability of an object, however, depends not only on its apparent
magnitude, but also on its morphology and mean SB: for a given apparent
magnitude, very extended, and hence low--SB, objects will have a lower
signal-to-noise than a compact source. In practice, any image presents a SB
limit beyond which the sample will be  incomplete.  For a given  flux limit,
the SB limit translates, therefore, into an upper limit on the size for which
an object can  still be detected. To determine the completeness of the FIRES
K--band image we created 100,000 artificial galaxies with intrinsic exponential
profiles and with structural parameter values covering the ranges
18$<$K$_s$$<$24, $0\farcs03<r_e<3\farcs0$ and
0$^\circ$$<$i$<$90$^\circ$\footnote{Galaxies with values of $n$ bigger than 1
are more centrally concentrated that an exponential and, hence, are easier to
detect at a given total magnitude. Therefore, our choice of $n=1$ is a
conservative one. We do note, however, that objects with $n<1$ will be harder
to detect at a given total magnitude; such objects are found to be dwarf
(faint) galaxies in the local universe and we assume that they will not be
observed in our high--z sample.}. The model images were randomly placed, 20 at
a time, into the K$_s$--band image, and  SExtractor was run using the same
parameters that were used to detect the real galaxies. Fig. \ref{ivofig} shows
the fraction of galaxies successfully detected by SExtractor at each input
value (K$_s$, r$_e$). Superimposed in Fig. \ref{ivofig}, we show the
K$_s$--band size and apparent magnitude for our sample objects. Also shown are
lines of constant central SB for exponential models (n=1). Even for the
conservative assumption of an exponential profile, we are complete over almost
the entire  range spanned by our sample galaxies.  This can be understood
simply because our NIR images are so deep.  Our sample selection threshold is
$\sim3$ magnitudes brighter than the 3-sigma detection limit.

 The exact SB limit for real distributions of galaxies is more complex, as
galaxies have a range of profile shapes, with different S\'ersic indices $n$
and hence different central SBs. Indeed, the data show no clearly defined
threshold. Nonetheless, as a conservative estimate of our completeness limit we
adopt a threshold at a central SB of $\mu_K$(0)=23.5 mag/arcsec$^2$, for which
we are 90\% complete for an exponential model. Objects with this  $\mu_K$(0)
that are more concentrated than an exponential would be detected with even more
completeness than 90\%. We have also found that, at the SB limit, we can
retrieve the sizes to within $\sim 20\%$ of objects with $n$=1. However, for
exponential objects near our SB limit, we underestimate the magnitude by a
median of 0.25 mag (and $>$0.4 mag for 25\% of these objects). This has to do
with the way SExtractor measures magnitudes, which depends on apparent SB. We
have also checked for possible incompleteness effects around the observed
K$_s$=23.5 magnitude limit, because of small systematic underestimates of
measured magnitudes, but find that they are not significant. The corrections to
total magnitudes for observed galaxies near the SB limit are, however, 
uncertain.  To be conservative we choose not to correct the flux at the SB
limit and note that the application of any correction for missed flux would
simply increased the derived luminosities of our galaxies.

\begin{figure}
\plotone{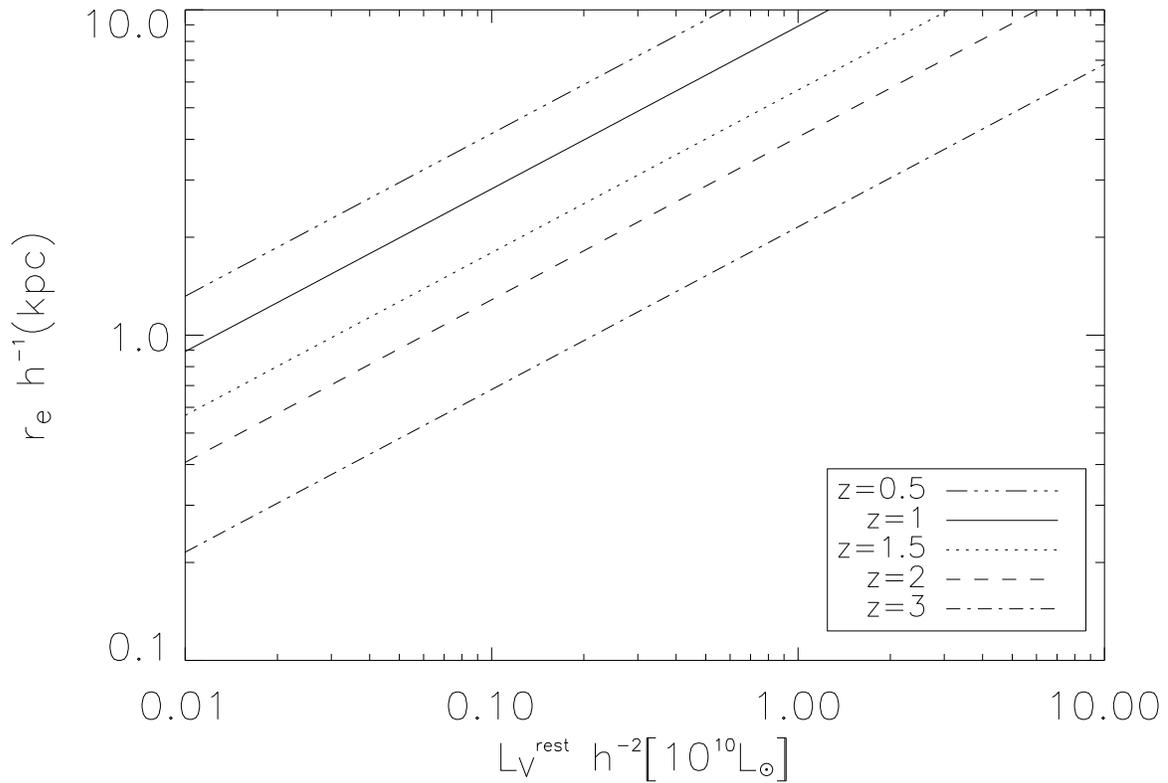}

\caption{The 90\% completeness tracks in effective radius for an exponential
 model with central surface brightness at K$_s$ of 23.5 mag/arcsec$^2$
 (see Fig.~\ref{ivofig}).  We adopt these conservative limits when
 creating mock high redshift catalogs of SDSS galaxies.  To convert
 from observed K-band magnitudes to rest--frame V--band luminosities
 we used an Scd template (see Fig. 11c from Rudnick et al. 2001).}

\label{relimit}

\end{figure}

    In Fig.~\ref{relimit} we show how our conservative SB limit translates into
 the 90\% completeness track in parameter plane of r$_e$ and L$_{V,rest}$. For
 a given redshift, we are less than 90\% complete for exponential galaxies with
 an effective radius larger than the corresponding line in Fig.~\ref{relimit}. 
 Due to (1+z)$^4$ SB dimming, redshift plays a very large role in this
 detectability. Similarly, for a given luminosity the maximal disk size to
 which we are complete will decrease with increasing redshift.

\section{ANALYSIS}

The selection effects  will affect the distribution of points in
 Fig.~\ref{rawdata} and make it impossible to read--off any size evolution, or
 lack thereof, without careful modelling. We explore evolutionary trends in the
 distribution of the galaxies in the above diagrams, by taking a z=0
 luminosity-size (and mass-size) relation and by drawing a mock high redshift
 catalogs from these relations, subject to the redshift--dependent selection
 effects.

\subsection{Simulating the local luminosity/mass vs. size relations at high
redshifts}

\begin{figure}
\plotone{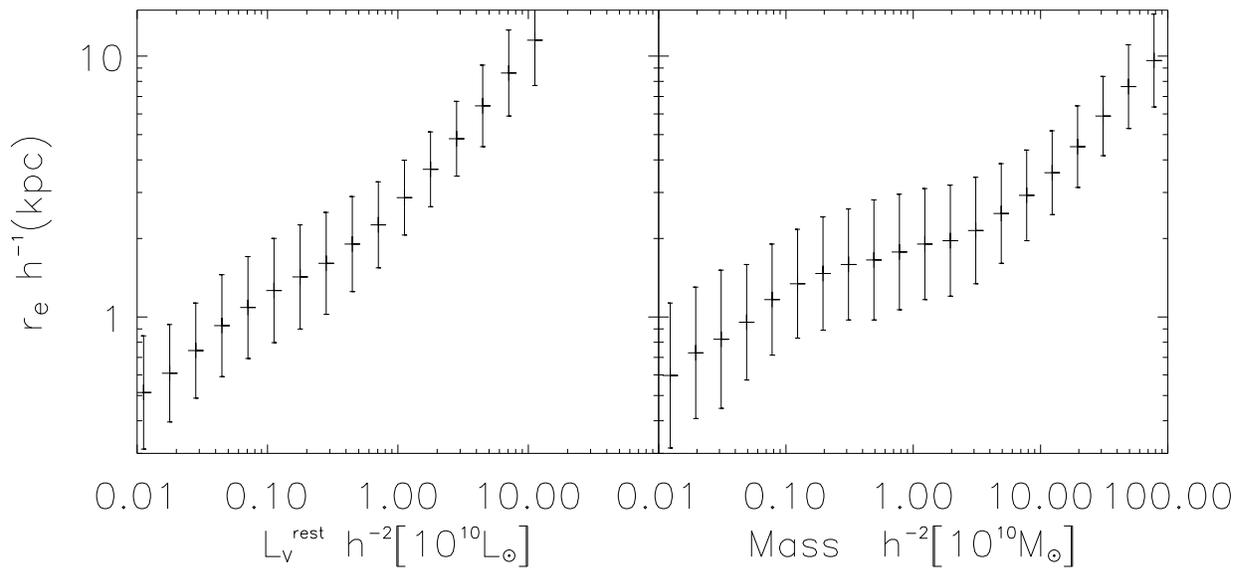}
\caption{$Left$: The median and dispersion of the distribution of the S\'ersic
half--light radius of the SDSS galaxies (in the g--band) as a function of the
g--band luminosity (the closest available filter to our V--band). $Right$:  Same
as before but as a function of the stellar mass. Note that the luminosity and
the mass extends in this figure up to 10$^{12}$ solar luminosities (masses). }

\label{figsdss}
\end{figure}

 The Sloan Digital Sky Survey (SDSS, York et al. 2000) is providing an
 unprecedented database of $\sim$ 10$^6$ nearby galaxies with spectroscopic
 redshifts and multi-band photometry. In particular, it has been used to derive
 the size distribution of present--epoch galaxies versus their luminosities and
 stellar masses (Shen et al. 2003). Shen et al. show the median and dispersion
 of the distribution of S\'ersic half--light radius (Blanton et al. 2003) for
 different bands as a function of the luminosity and of the stellar mass. We
 have used their g--band (the closest available filter to our V--band) size
 distributions (S.  Shen, 2003, private communication) as a local reference of
 the size distribution of galaxies in the nearby universe. We note that whereas
 Shen et al. show separately the distribution of early and late--type galaxies,
 we use their combination of these two subsamples into one, to make a direct
 comparison with our sample. For nearby galaxies of a given luminosity, Shen et
 al. propose  the following size log--normal distribution
 with median $\bar{r}_e$ and logarithmic dispersion $\sigma$:

\begin{equation}
f(r_e|\bar{r}_e(L),\sigma(L))=\frac{1}{\sqrt{2\pi}\sigma(L)}
 e^{-\frac{\ln^2 (r_e/\bar{r}_e(L))}{2\sigma^2(L)}}\frac{dr_e}{r_e},
\label{sdss}
\end{equation}
illustrated in Fig. \ref{figsdss}.

 The SDSS relations are used to test the null hypothesis, i.e., that the
 luminosity--size or mass--size relations do not change with redshift. It is
 important to note that this does not imply that the galaxies themselves do not
 change; they could certainly evolve `along' such a relation. To test the null
 hypothesis, we construct distributions of SDSS galaxies as they would be
 observed at high redshift, mimicking our observations as follows: every
 simulated distribution of galaxies contains the same number of objects than
 our FIRES sample (i.e. 168). We pick a luminosity and redshift pairs at random
 from our observed sample. For this luminosity L, we evaluate a size at random
 from the local size sample distribution provided by the SDSS data (Eq.
 \ref{sdss}), by solving the following implicit equation:

\begin{equation}
F(r_e|L)=\frac{1}{2}\bigg\{1-\Phi\bigg(\frac{\ln
(r_e/\bar{r}_e(L))}{\sqrt{2}\sigma(L)}\bigg)\bigg\};
\label{sdss2}
\end{equation}
where $\Phi$ is the error function and F(r$_e|$L) is randomly distributed in [0,1].

 For every effective radius  drawn from Eq. \ref{sdss2} we analyze if this
 galaxy (characterized by r$_e$, L, z) would be observed within the
 completeness limit of Fig. \ref{relimit}. If it is larger, it is not taken
 into account in our simulated distribution. The process for selecting a new
 galaxy is repeated until we have a mock sample with the same number of objects
 as galaxies observed.

 This procedures assures that the simulated galaxy distribution follows the
 same redshift and luminosity distribution as the observed sample and also is
 affected by the same selection effects. An analogous procedure is repeated in
 the case of the size--mass relation replacing L with M$_\star$ in Eqs.
 \ref{sdss} and \ref{sdss2}\footnote{Shen et al. use also a log--normal
 distribution for the size--mass relation. Their mass evaluation rests in the
 Kroupa (2001) IMF (Initial Mass Function). The stellar mass modelling of the
 FIRES data, however, uses a Salpeter IMF. In our simulations we have followed
 the procedures suggested in Kauffmann et al. (2003) of using
 Mass$_{IMF,Salpeter}$=2$\times$Mass$_{IMF,Kroupa}$ to transform from SDSS data
 to our data.}. At this time, to account for the selection effects, we select
 a  (M/L, L, z) triplet at random  from the observed distribution.

\begin{figure}

\plotone{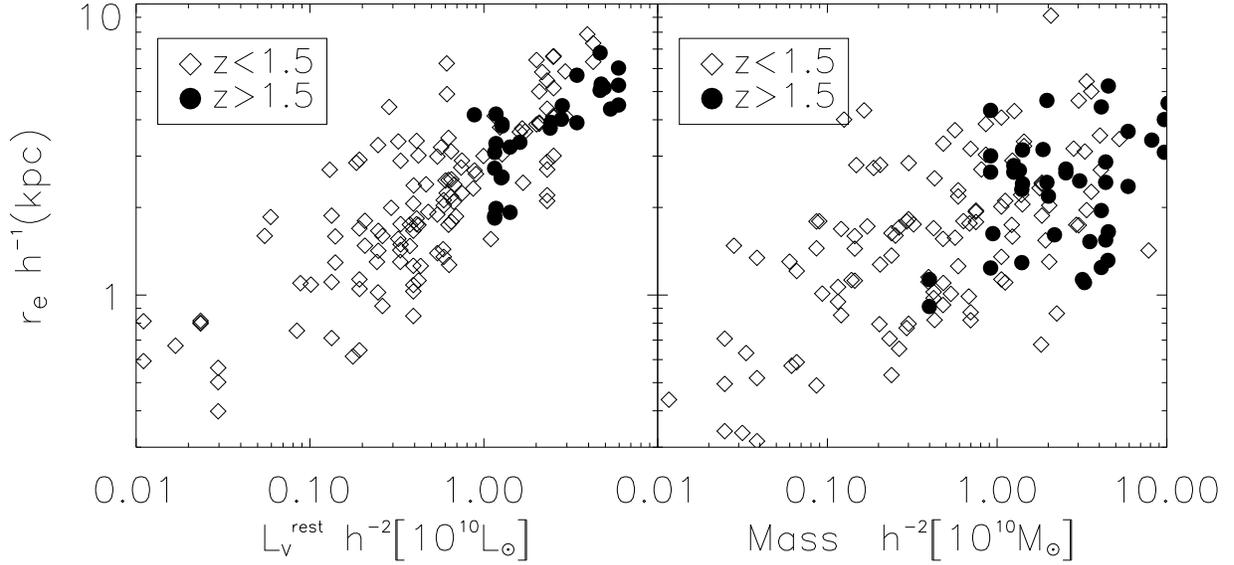}

\caption{Predictions of the null hypothesis: $Left$: Simulated distribution of
rest--frame optical sizes  versus the rest--frame V--band luminosities  for the
SDSS data is shown (see text for details). Galaxies with redshifts smaller than
1.5 are represented with open diamonds  and galaxies with redshifts bigger than
1.5 with solid circles. $Right$: Same as before but with the stellar mass.}

\label{simulsdss}

\end{figure}

 Fig. \ref{simulsdss} shows an example of how the SDSS galaxies would be
 distributed in the size diagrams with the same luminosity, mass and redshift
 distribution as the galaxies observed in FIRES. A comparison of Fig.
 \ref{rawdata} and \ref{simulsdss} shows that the simulated SDSS data have a
 luminosity--size and a mass-size relation that is tighter than the  observed
 FIRES relations. If the luminosity--size  relation from SDSS remained
 unchanged with increasing z, we would not expect to find small and luminous
 objects at high redshift, however these objects are present in the FIRES
 observations (see Fig. \ref{rawdata}).

 To quantify if these qualitative differences between the observed
 distributions and the simulated null hypothesis are significant, we ran the
 generalization of the Kolmogorov--Smirnov (K--S) test to two--dimensional
 distributions (Fasano \& Franceschini 1987). We create 1000  SDSS realizations
 (both for luminosity and mass). For all the simulations the rejection
 probability is bigger than 99.9\%. Consequently, we conclude that neither the
 luminosity--size nor the mass--size observed relations satisfy the null
 hypothesis.

 To account for measurement errors in the size estimates of the FIRES galaxies,
 we also create mock ``FIRES'' data point distributions. To create these
 distributions we randomly vary  every observed effective radius using a normal
 distribution with standard deviation equal to the size error measured for each
 galaxy. We make 1000 mock ``FIRES'' and compare for each  the rejection
 probability  between the SDSS and the mock ``FIRES'' data. The rejection
 probability for all mock samples is again more than 99.9\% in the luminosity
 and the mass relations. So, the intrinsic dispersion of our measurements are
 unable to explain the difference with the SDSS simulated data.


We also explored the sensibility of the adopted FIRES selection limits through
 simulations: we have evaluated the SDSS DF including different central SB
 limit ranging from 23 to no restrictions at all. We do not find any
 significant difference in our results. As we expected due to the depth of our
 images,  uncertainties in the selection effects do not affect our analysis.

\subsection{Testing the hypothesis of evolution}

 The no--evolution hypothesis, that the size relations for all galaxies are
redshift independent, can be rejected both for the L$_V$--r$_e$ and
M$_\star$--r$_e$ relations. To quantify and constrain the evolution of these
relations with redshift, we need to devise an evolution model. In the absence
of clear--cut theoretical predictions, we have resorted to a heuristic
parameterization that draws on the observed local distribution. We have assumed
that the  log--normal size distribution  (Eq. \ref{sdss}) applies at all
redshifts, but with evolving parameters: \begin{eqnarray}
\bar{r}_e(L,z)=\bar{r}_e(L,0)(1+z)^{-\alpha} \label{sdss3} \\
\sigma(L,z)=\sigma(L,0)(1+z)^{\beta}. \label{sdss4}\end{eqnarray} Here,
$\bar{r}_e(L,0)$ and $\sigma(L,0)$ are the median size and dispersion provided
at z=0 by the Shen et al. (2003) data, and $\alpha$ and $\beta$ describe the
redshift evolution. Note that Eq. \ref{sdss3} and Eq \ref{sdss4} imply the same
size evolution for all the galaxies independently of their luminosity. We also
assume an analogous parameterization for the masses. Both for L$_V$--r$_e$ and
M$_\star$--r$_e$ we explore the ranges between [-2,3] for $\alpha$ and between
[-2,2] for $\beta$.

 As for the null hypothesis, we generate simulated galaxy distributions for
every pair ($\alpha$, $\beta$) and we ran K--S test between these simulations
and the observed data. Neither for L$_V$--r$_e$ nor for M$_\star$--r$_e$  could
we  produce evolutionary scenarios ($\alpha$, $\beta$) whose distribution
functions are in agreement with what we see in the FIRES data.  However, one
must bear in mind that not all the luminosities can be observed over the full
redshift range  (see Fig. \ref{zphotlum}). To understand better what possible
evolution our data imply and to avoid luminosity dependent redshift ranges we
decided to create more homogeneous subsamples by splitting our sample in three
different luminosity (mass) bins. Both our luminosity--size and mass--size
observed distributions have been divided in three luminosity (mass) bins as
detailed in Table \ref{tabbin}\footnote{The lowest luminosity (mass) galaxies
with L$<$0.3$\times$10$^{10}$ L$_\odot$ (M$_\star$$<$0.3$\times$10$^{10}$
M$_\odot$) are not presented in this analysis. These galaxies have z$\lesssim$1
and, consequently, the results coming from this subsample are largely affected
by the cosmic variance associated with the small volume enclosed in the HDF-S
over this redshift range.}. The splitting of our sample into these intervals
avoids that low luminosity (mass) galaxies at low redshift dominate the results
of our analysis. The  mean redshifts for the low/intermediate/high luminosity
(mass) bins are 1.0, 1.6 and  2.0, respectively. These different sets measure
evolution in a different luminosity (mass) range and a different redshift
range, and splitting them helps to make this clear.


We now can check whether the observed FIRES relations can be explained if the
evolutionary parameters $\alpha$ and $\beta$ depend on the luminosity (mass),
and write out the new distribution function explicitly,   combining Eqs.
\ref{sdss3} and \ref{sdss4} with Eq. \ref{sdss}:

\begin{equation}
g(r_e|\bar{r}_e,\sigma,z)=\frac{1}{\sqrt{2\pi}\sigma}
\frac{dr_e}{r_e}\frac{dz}{(1+z)^{\beta}} e^{-\frac{\ln^2
(r_e/\bar{r}_e(1+z)^{\alpha})}{2\sigma^2(1+z)^{2\beta}}}  \label{fires}
\end{equation}

\begin{figure}

\plotone{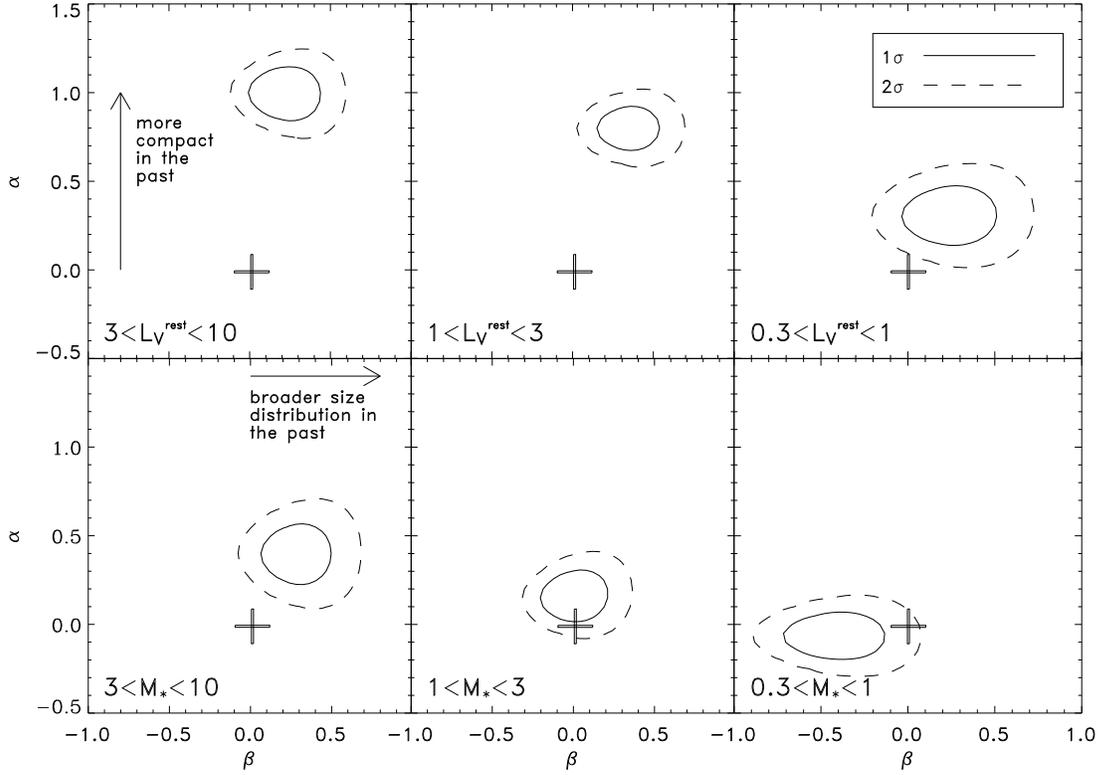}
\vspace{1cm}

\caption{Likelihood contours representing the evolution in size and dispersion
in the $\alpha$--$\beta$ plane. The top (bottom) row shows the evolution of the
luminosity--size (mass--size) relation. Solid line represents the 1$\sigma$
confidence level contour, dashed line the 2$\sigma$ confidence level. The cross
shows the position of no--evolution in this plane.  Positive values of $\alpha$
represent decreasing values of the size with redshift. Positive values of
$\beta$ represent increasing the instrinsic dispersion $\sigma$ of the
population with redshift. Both the luminosity and the mass are in units of
10$^{10}$ solar value. Both the no evolution model and a luminosity independent
evolution model are less likely than the luminosity dependent model.}

\label{likeli}
\end{figure}

The expression in Eq. \ref{fires} is a probability density and we use this to
evaluate $\alpha$ and $\beta$ using a Maximum Likelihood Method.  For each
luminosity/mass subsample  we show in Fig. \ref{likeli} the likelihood contours
(1$\sigma$ and 2$\sigma$) in the plane of $\alpha$ and $\beta$.  We have also
included as a reference the point $\alpha$=0 and $\beta$=0 which indicates the
case of no--evolution at this plane. The top row shows the evolution of the
size distribution at a given luminosity. The mean size at a given luminosity
changes significantly for luminous galaxies: at 
$<$L$_V$$>$$\sim$2$\times$10$^{10}$h$^{-2}$L$_\odot$ galaxies were typically
three times smaller at z$\sim$2.5 than now, and 4 times smaller for
L$_V$$>$3$\times$10$^{10}$h$^{-2}$L$_\odot$. A luminosity independent model
($\alpha$ and $\beta$ independent of L) is less likely than our luminosity
dependent model, justifying our choice of three subsamples. For the sizes at a
given stellar mass the picture is qualitatively different: there is no evidence
for significant evolution of the r$_e$--M$_\star$ relation, except for the most
massive bin, M$_\star$$>$3$\times$10$^{10}$h$^{-2}$M$_\odot$. At
$<$M$_\star$$>$$\sim$2$\times$10$^{10}$h$^{-2}$M$_\odot$ the relation may only
have changed by 20\% since z$\sim$2.5. The implied size evolution at $z$=2.5 in
each luminosity (mass) bin is summarized in Table \ref{tabbin}. Note that in
all cases the evidence for an evolving scatter of the L$_V$,M$_\star$--r$_e$
relations is marginal.

\begin{figure}
\plotone{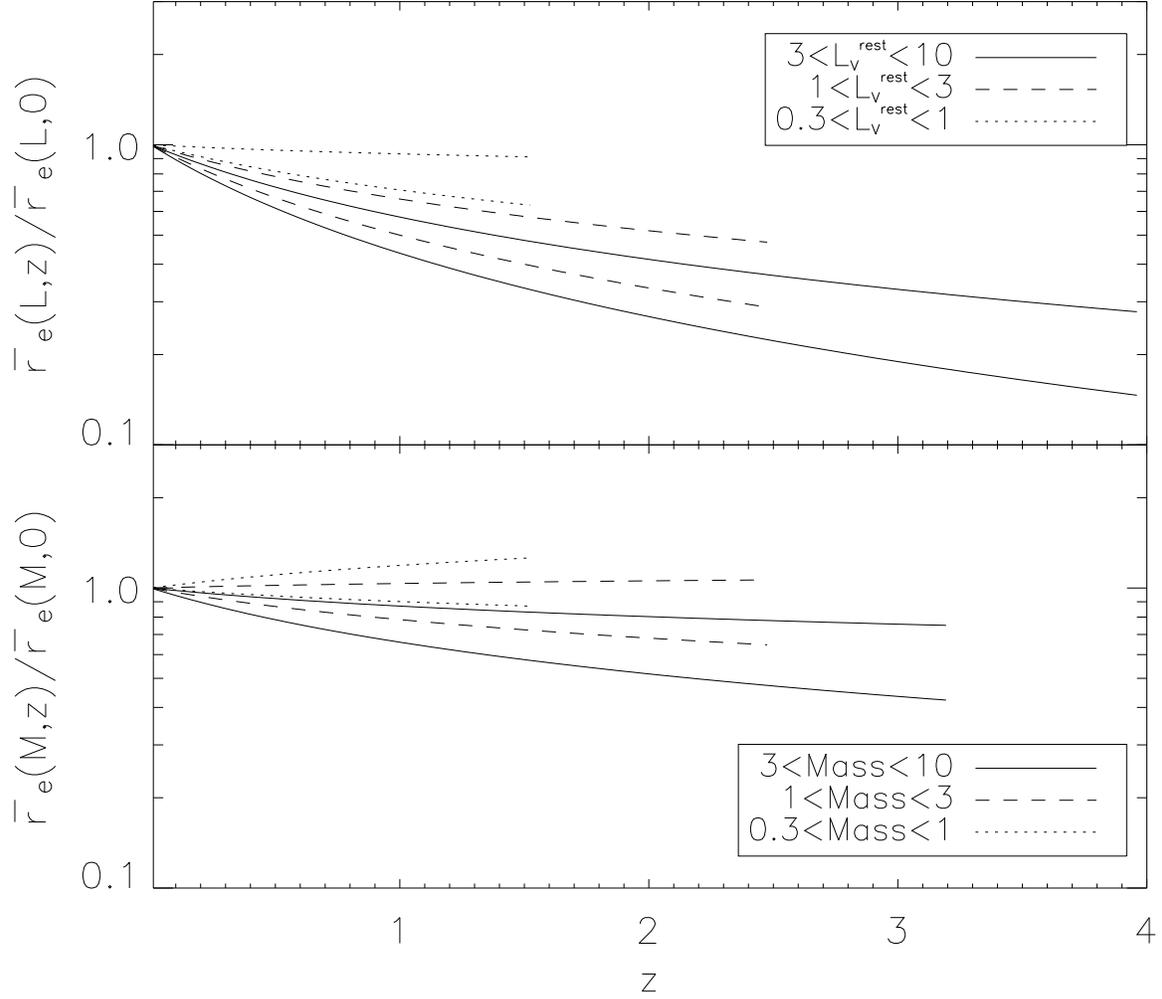}
\vspace{1cm}
\caption{The  ratio between the SDSS mean size and the expected
mean size as a function of z  is shown both for the luminosity
and for mass relations. Both the luminosity and the mass are in units of
10$^{10}$
solar value. Equal style lines enclose the 1$\sigma$ variation.}
\label{qratio}
\end{figure}

 In Fig. \ref{qratio}, we visualize these results in a different way: we show
 the ratio between the present epoch mean size for every luminosity (mass) bin
 and the expected mean size as a function of redshift using the  $\alpha$
 values derived from the FIRES data (i.e. we show
 $\bar{r}$$_e(L,z)$/$\bar{r}$$_e(L,0)$=(1+z)$^{-\alpha}$). The same is done for
 the mass. This figure shows the region enclosed by the 1$\sigma$ level
 confidence contours. The lines stop at the limiting redshift z$'$ we are able
 to explore for the different luminosity (mass) bins.

 Again, Fig. \ref{qratio} shows that high--z galaxies (most luminous bin) at
 z$\sim$2.5 are more compact (a factor of 4) than the nearby equally luminous
 galaxies. On the other hand, high--z galaxies differ only slightly in size at
 a given mass  from the present--epoch. In the middle luminosity (mass) bin the
 evolution with z appears to be less important. For the L$_V$--r$_e$ relations
 the dispersion of the high--z population increases in all the cases. This
 increase is, however, relatively moderate (a factor of 1.2--2.). We discuss
 how we can understand these results in the following section.

\section{DISCUSSION}

Using the observed nearby SDSS size relations (Shen et al. 2003) as the correct
local references, the observed FIRES size--luminosity and  mass--size
distributions at high--z show a very different degree of evolution. The
mass--size relation has remained practically  unchanged whereas, the
size--luminosity has  evolved significantly: there are many more compact
luminous objects at high--z than now. How can we re--concile these two
observational facts?


In absence of M/L evolution with time, a change in the size--luminosity
relation with z would imply the same degree of evolution in the mass--size
relation. However, the mean M/L ratio decreases with increasing z. In the
nearby universe, most galaxies have large M/L ratios (see Fig. 14 of Kauffmann
et al. 2003). In contrast, FIRES galaxies at z$>$2, at  all luminosities, have
M/L ratios of the order $\sim$1 (Rudnick et al.  2003b). Consequently, although
we observe a strong evolution in the luminosity--size relation, the decrease of
M/L avoids a significant change in the observed mass--size relation.


 We can therefore characterize our observed high--z galaxy population as
follows: small to medium size objects (effective radius $\sim$1.5 h$^{-1}$ kpc)
not very massive ($\sim$3$\times$10$^{10}$h$^{-2}$M$_\odot$) but often very
luminous ($\sim$3$\times$10$^{10}$h$^{-2}$L$_\odot$) in the V--band. The above
picture does not mean that large galaxies can not be found at high z (Labb\'e
et al. 2003b), but that they are relatively rare.

 Traditionally the high--z population has been selected by the Lyman Break
 selection technique (Steidel \& Hamilton 1993) known to select luminous
 unobscured star--forming galaxies. However, dust obscured  or UV--faint
 galaxies may have been missed. The galaxies in FIRES are selected from very
 deep near infrared K$_s$--band imaging, and consequently, are expected to be
 less affected by this problem and give a more complete mass census of the
 high--z universe. In fact, the population of galaxies under study consists in
 part of a red population (Franx et al. 2003), which would be largely missed by
 the Lyman Break technique, but whose volume number density is estimated to be
 half that of LBGs at z$\sim$3.

 Hierarchical models for structure formation in a $\Lambda$ dominated universe
 predict that LBGs have typical half--light radii of $\sim$2 h$^{-1}$ kpc (Mo,
 Mao \& White 1999) in good agrement to the size of the galaxies we are
 measuring and to the observed sizes for LBGs of other authors (Giavalisco,
 Steidel \& Macchetto 1996, Lowenthal et al. 1997). Interestingly, other
 authors have observed LBGs using optical filters and, consequently, these
 sizes are UV sizes. The fact that their measures and ours (which are in the
 optical rest--frame) do not differ significantly could be evidence that the
 star formation of the LBGs is extended over the whole object\footnote{There is
 some evidence that the LBGs morphology depends not much on the wavelength,
 remaining essentially unchanged from the far-UV to the optical window
 (Giavalisco 2002 and references therein).}. In fact, if we select in our
 sample those galaxies with L$_V$$>$2$\times$10$^{10}$h$^{-2}$L$_\odot$ and
 z$>$2.5, 1/2 of this subsample would be considered as LBGs following the Madau
 et al. (1996) color criteria.  We will explore the relation of UV and optical
 sizes in a forthcoming paper.

\begin{figure}
\plotone{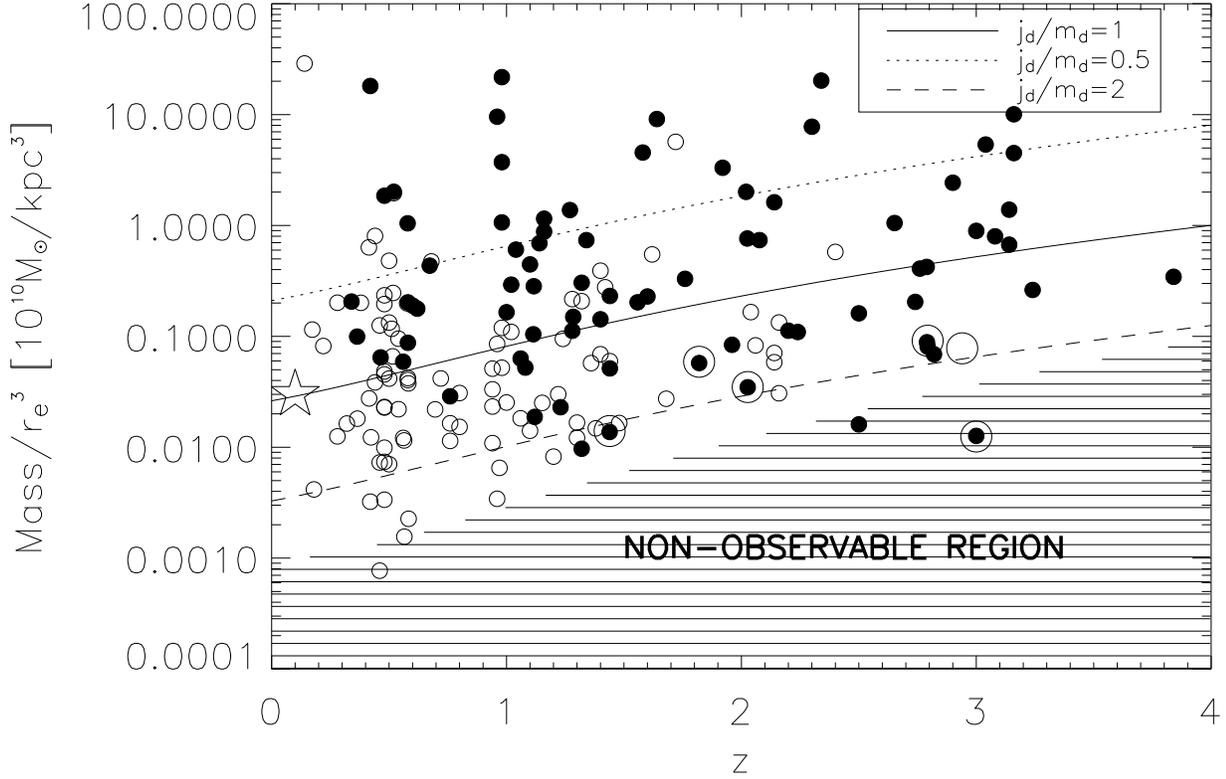}
\vspace{1cm}

\caption{The internal density of the observed galaxies versus $z$. In order to
make a direct comparison with the theory expectation we have used $h$=0.7.
Observed galaxies are separated in two groups: solid points represent galaxies
more massive than 10$^{10}$M$_{\odot}$. These galaxies are observable over the
full $z$ range. The expected internal density M/r$_e^3$ of disk galaxies  just
formed at each redshift  according to the Mo, Mau \& White (1998) model is
shown for three different specific angular momentum $j_d/m_d$ values. We have
used m$_d$=0.05 and $\lambda$=0.05. The internal density of the Milky Way
galaxy (star) is shown for comparison. Also, we have encircled those galaxies
proposed to be large disk--like galaxies at high--z following Labb\'e et al.
(2003). Note that there is a galaxy in the Labb\'e et al. analysis which is not
in our sample because it is outside of the field of view selected in this
study.}

\label{momau}
\end{figure}

``High--redshift disks are predicted to be small and dense, and could plausibly
merge together to form the observed population of elliptical galaxies'' (Mo,
Mau \& White 1998). We have made a simplistic comparison between the above
prediction and our data in Fig. \ref{momau}. The lines represent the expected
internal mass density M/r$_e^3$ of disks galaxies just formed at each redshift
for three different values of the specific angular momentum. These lines are
evaluated combining the Eqs. (4) and (12) of Mo et al. (1998) and assume a
constant fraction of the mass of the halo which settles into the disk,
m$_d$=0.05, and a constant spin parameter of the halo $\lambda$=0.05. With
these two assumptions the internal mass density  increases with $z$ as the
square of the Hubble constant $H(z)$.

 Galaxies more massive than 10$^{10}$M$_\odot$ are observable over the complete
range in redshift. The measured mean internal density for this galaxy
population appears to evolve only slightly with $z$, in agreement with the lack
of strong evolution in the size--mass relation. If all the galaxies present in
Fig. \ref{momau} were disks, their distribution would not be compatible with
the theoretical expectation. However, we must take into account that we are
observing a mix of galaxy types and not only disk galaxies. In order to address
this point we have made a visual galaxy classification of all the objects with
z$<$1.5 and mass larger than 10$^{10}$M$_\odot$. We can do that because of the
high--resolution of the HST images in the I$_{814}$--band filter. This
examination showed that the dense objects
(M/r$_e^3>$10$^{10}$M$_\odot$/kpc$^3$) with z$<$1.5 appear to be all
ellipticals and that the late type fraction appears to increase as one moves to
lower density objects. Unfortunately, we cannot make a similar analysis for
z$>$1.5 and the question of whether the high density objects we observe there
are disk dominated remains unsolved. However, it is highly tempting to propose
that our high $z$ dense population could be the progenitors of the nearby dense
elliptical galaxies.



Independently of the nature of the LBGs  and of the red population it is clear
that in order to reach the mass and sizes of the nearby galaxies an
evolutionary process must be acting on the high--z population. Recently, Shen
et al. have proposed, for the early--type galaxies, a simple model of mass and
size evolution based on subsequent major mergers. This model  explains the
observed relations for these kind of galaxies  between mass and size in the
SDSS data, i.e. R$\propto$M$^{0.56}$. In the Shen et al. picture  two galaxies
with the same mass (M$_1$=M$_2$) and radius (R$_1$=R$_2$)  merge forming a new
galaxy with mass M=2M$_1$ and radius R=$2^{0.56}$R$_1$. If this process is
repeated $p$ times: M=2$^p$M$_1$ and R=$2^{0.56p}$R$_1$. If we take a galaxy at
$z$=3 (following our mass and size estimates) with
M$_1$=3$\times$10$^{10}$h$^{-2}$M$_\odot$ and R$_1$=1.5h$^{-1}$ kpc this
implies that after 3 major mergers M=24$\times$10$^{10}$h$^{-2}$M$_\odot$ and
R=4.8 h$^{-1}$ kpc, in excellent agreement to the values we see in nearby
galaxies (M=24$\times$10$^{10}$h$^{-2}$M$_\odot$ and
R=5 h$^{-1}$ kpc, see Fig. \ref{figsdss}). These numbers  may suggest that the massive and dense  high--z
population can be understood as the progenitors of the bright and massive
nearby early type galaxies.

\section{SUMMARY}

Using ultra--deep near--infrared images of the HDF-S we have analyzed the
rest--frame optical band sizes of a  sample of galaxies selected down to
K$_s$=23.5. This has  allowed us to measure the evolution of the
luminosity--size and mass--size relation out to z$\sim$3. This is the first
time that the rest--frame V--band sizes of such distant galaxies have been
systematically analyzed as a function of stellar luminosity and stellar mass.

We compared  our observed luminosity--size and mass--size relations to those
measured in the nearby universe by the SDSS data (Shen et al. 2003). For this
comparison we have analyzed in detail the detectability effects that high--z
observations impose on the observed relations. From this comparison, assumming
the Shen et al. distributions are correct, we found:

\begin{enumerate} \item The size--luminosity relation has evolved since
z$\sim$2.5. Luminous objects (L$_V$$\sim$3$\times$10$^{10}$h$^{-2}$L$_\odot$)
at z$\sim$2.5 are 4 times smaller than  equally luminous galaxies today. \item
The size--stellar mass relation has remained nearly constant since z$\sim$2.5:
for $<$M$_\star$$>$$\sim$2$\times$10$^{10}$h$^{-2}$M$_\odot$ the change is 
20($\pm$20)\%;  for stellar masses larger than
3$\times$10$^{10}$h$^{-2}$M$_\odot$ the characteristic mean size change is
40($\pm$15)\%. \end{enumerate}

The above results are reconciled by the fact  the M/L values of high--z
galaxies are lower than nowadays $<$M/L$>$$\sim$1 (Rudnick et al. 2003b).
Consequently, the brightest high--z galaxies are a group composed of a high
internal luminosity density population but with a mean internal stellar mass
density not much higher than found in the nearby universe. The observed small
sizes of distant galaxies found here and in previous studies for LBGs
(Giavalisco, Steidel \& Maccheto 1996; Lowenthal et al.  1997; Ferguson et al.
2003) are in agreement with the small evolution of the mass--size relation
because the typical masses of $z$=3 galaxies are substantially smaller than
those at low redshift.


\acknowledgements

We are happy to thank Shiyin Shen for providing us with the Sloan Digital Sky
Survey data used in this paper and Boris H\"au\ss ler for running GALFIT. We
thank the staff at ESO for the assistance in obtaining the FIRES data and the
the Lorentz Center for its hospitality and support. We also thank the anonymous
referee for useful comments.

Funding for the creation and distribution of the SDSS Archive has been provided
by the Alfred P. Sloan Foundation, the Participating Institutions, the National
Aeronautics and Space Administration, the National Science Foundation, the US
Department of Energy, the Japanese Monbukagakusho, and the Max-Planck Society.
The SDSS Web site is http://www.sdss.org. The SDSS is managed by the
Astrophysical Research Consortium (ARC) for the Participating Institutions. The
Participating Institutions are the University of Chicago, Fermilab, the
Institute for Advanced Study, the Japan Participation Group, Johns Hopkins
University, Los Alamos National Laboratory, the Max-Planck-Institut f\"ur
Astronomie (MPIA), the Max-Planck-Institut f\"ur Astrophysik (MPA), New Mexico
State University, University of Pittsburgh, Princeton University, the US Naval
Observatory, and the University of Washington.

\clearpage

\clearpage
\begin{deluxetable}{cccccccccc}

\tablecolumns{10}
\tablewidth{0pc}
\tablehead{
Galaxy & X & Y & K$_{s,tot}$ & r$_e$ & $\epsilon$ & L$_V$(10$^{10}$ h$^{-2}$ L$_\odot$)   &
M(10$^{10}$ h$^{-2}$ M$_\odot$)  & $z$ & Filter}

\startdata
     793 &  3538.3 & 3496.9 & 21.33  &   0.04 & 0.10  &    0.01 &  0.01  &     0.140   & I$_{814}$ \\
     244 &  2272.6 & 1587.5 & 20.71  &   0.40 & 0.67  &    0.06 &  0.09  &     0.173$^a$    & I$_{814}$ \\
     283 &  2063.5 & 1660.1 & 23.34  &   0.51 & 0.62  &    0.01 &  0.01  &    0.180   & I$_{814}$ \\
     314 &  2741.5 & 1815.3 & 21.61  &   0.24 & 0.08  &    0.02 &  0.02  &   0.220   & I$_{814}$ \\
     288 &   501.6 & 1633.9 & 23.23  &   0.39 & 0.09  &    0.01 &  0.02  &     0.280   & I$_{814}$ \\
     464 &   943.0 & 2017.5 & 20.76  &   0.30 & 0.44  &    0.11 &  0.20  &    0.280   & I$_{814}$ \\
     227 &  1393.0 & 1430.0 & 21.51  &   0.60 & 0.35  &    0.12 &  0.17  &    0.320   & I$_{814}$ \\
     184 &   626.9 & 1387.0 & 18.89  &   0.55 & 0.21  &    0.75 &  1.92  &   0.340$^a$   & I$_{814}$ \\
     446 &  2417.6 & 2080.2 & 20.05  &   0.56 & 0.36  &    0.39 &  1.11  &   0.364$^a$   & I$_{814}$ \\
     528 &  2478.4 & 3018.5 & 20.53  &   0.71 & 0.64  &    0.28 &  0.42  &   0.365$^a$   & I$_{814}$ \\
      53 &  2954.6 &  522.7 & 23.26  &   0.14 & 0.23  &    0.02 &  0.03  &     0.380   & I$_{814}$ \\
     663 &  3106.0 & 3832.1 & 21.28  &   0.17 & 0.08  &    0.19 &  0.26  &    0.414$^a$   & I$_{814}$ \\
     763 &  1258.2 & 3079.6 & 19.89  &   0.76 & 0.15  &    0.56 &  0.96  &    0.415$^a$   & I$_{814}$ \\
     138 &  3709.8 & 1029.9 & 19.83  &   0.14 & 0.40  &    0.48 &  4.08  &    0.420   & I$_{814}$ \\
     189 &  1678.4 & 1190.2 & 23.00  &   0.64 & 0.56  &    0.06 &  0.07  &     0.420   & I$_{814}$ \\
     256 &  1619.2 & 1632.0 & 21.84  &   0.54 & 0.26  &    0.17 &  0.16  &    0.423$^a$   & I$_{814}$ \\
     307 &  3037.7 & 1762.6 & 21.79  &   0.43 & 0.29  &    0.15 &  0.27  &   0.440   & I$_{814}$ \\
     374 &  1205.1 & 2466.3 & 22.80  &   0.13 & 0.22  &    0.03 &  0.16  &    0.440   & I$_{814}$ \\
      29 &  3817.2 &  486.8 & 20.58  &   0.34 & 0.24  &    0.64 &  0.48  &     0.460   & I$_{814}$ \\
      75 &  2162.5 &  727.1 & 23.14  &   0.94 & 0.05  &    0.08 &  0.06  &     0.460   & I$_{814}$ \\
     708 &  2656.4 & 3737.6 & 21.27  &   0.75 & 0.20  &    0.35 &  0.30  &    0.460   & I$_{814}$ \\
     292 &   781.1 & 1876.3 & 18.88  &   0.86 & 0.48  &    1.15 &  4.03  &   0.464$^a$   & I$_{814}$ \\
      87 &   459.7 &  700.5 & 23.26  &   0.29 & 0.30  &    0.05 &  0.05  &     0.480   & I$_{814}$ \\
      95 &  3873.2 &  715.5 & 23.33  &   0.11 & 0.30  &    0.05 &  0.03  &     0.480   & I$_{814}$ \\
     344 &  3579.9 & 2599.2 & 22.78  &   0.42 & 0.50  &    0.10 &  0.07  &     0.480   & I$_{814}$ \\
     415 &  2401.5 & 2278.5 & 21.34  &   0.30 & 0.21  &    0.18 &  0.56  &   0.480   & I$_{814}$ \\
     442 &  3236.2 & 2130.7 & 22.46  &   0.54 & 0.34  &    0.12 &  0.11  &    0.480   & I$_{814}$ \\
     459 &  3410.0 & 2059.1 & 23.14  &   0.19 & 0.17  &    0.08 &  0.03  &     0.480   & I$_{814}$ \\
     544 &  1405.0 & 2923.0 & 21.51  &   0.51 & 0.63  &    0.16 &  0.32  &   0.480   & I$_{814}$ \\
     673 &  2131.5 & 3486.8 & 22.87  &   0.24 & 0.49  &    0.09 &  0.06  &     0.480   & I$_{814}$ \\
     758 &  2271.3 & 3535.7 & 20.12  &   0.24 & 0.23  &    0.47 &  2.57  &   0.480   & I$_{814}$ \\
     766 &  2035.5 & 3411.7 & 22.13  &   0.73 & 0.53  &    0.19 &  0.13  &     0.480   & I$_{814}$ \\
      41 &  2006.4 &  560.3 & 21.32  &   0.27 & 0.33  &    0.33 &  0.29  &    0.500   & I$_{814}$ \\
     270 &  2703.9 & 1586.5 & 21.93  &   0.15 & 0.29  &    0.20 &  0.18  &   0.500   & I$_{814}$ \\
     345 &  3288.8 & 2575.0 & 21.74  &   0.71 & 0.00  &    0.23 &  0.28  &     0.500   & I$_{814}$ \\
     476 &  2056.1 & 1976.4 & 22.35  &   0.32 & 0.18  &    0.15 &  0.14  &     0.500   & I$_{814}$ \\
      91 &  2719.3 &  841.1 & 20.92  &   0.35 & 0.36  &    0.34 &  0.58  &     0.511$^a$    & I$_{814}$ \\
     339 &  2398.7 & 2627.2 & 21.54  &   0.32 & 0.31  &    0.36 &  0.26  &    0.515$^a$   & I$_{814}$ \\
     609 &  1716.1 & 3098.5 & 20.94  &   0.26 & 0.36  &    0.40 &  0.53  &    0.517$^a$   & I$_{814}$ \\
     247 &  4006.2 & 1587.3 & 20.35  &   0.17 & 0.16  &    0.67 &  1.26  &   0.520   & I$_{814}$ \\
     317 &  1837.8 & 1846.0 & 22.16  &   0.08 & 0.33  &    0.17 &  0.12  &     0.520   & I$_{814}$ \\
     482 &  1264.5 & 1950.4 & 22.62  &   0.21 & 0.12  &    0.13 &  0.12  &     0.540$^a$    & I$_{814}$ \\
     577 &  2577.1 & 2942.7 & 22.93  &   0.32 & 0.56  &    0.08 &  0.08  &	 0.540   & I$_{814}$ \\
      50 &  3581.5 &  687.1 & 20.11  &   0.54 & 0.29  &    0.88 &  1.22  &   0.560$^a$   & I$_{814}$ \\
     215 &  3602.8 & 1301.6 & 23.12  &   0.42 & 0.43  &    0.07 &  0.11  &   0.560	 & I$_{814}$ \\
      99 &  2736.9 &  923.1 & 20.93  &   0.72 & 0.62  &    0.42 &  0.56  &     0.564$^a$    & I$_{814}$ \\
     223 &  2446.8 & 1476.4 & 20.98  &   1.29 & 0.16  &    0.62 &  0.45  &    0.565$^a$   & I$_{814}$ \\
     234 &  2859.3 & 1472.7 & 20.28  &   0.33 & 0.26  &    0.51 &  0.96  &     0.577$^a$    & I$_{814}$ \\
     549 &  2982.7 & 2839.3 & 19.51  &   0.30 & 0.15  &    1.54 &  4.11  &    0.579$^a$   & I$_{814}$ \\
     153 &  3459.5 & 1028.6 & 22.46  &   0.31 & 0.67  &    0.15 &  0.16  &    0.580   & I$_{814}$ \\
     193 &  3572.2 & 1508.4 & 18.11  &   0.73 & 0.34  &    4.64 &  11.0  &     0.580$^a$    & I$_{814}$ \\
     409 &  1750.0 & 2270.8 & 19.46  &   0.55 & 0.09  &    1.72 &  2.07  &   0.580$^a$   & I$_{814}$ \\
     770 &  3363.7 & 3511.1 & 22.00  &   0.34 & 0.61  &    0.20 &  0.23  &    0.580$^a$   & I$_{814}$ \\
     410 &  1808.7 & 2284.3 & 22.01  &   0.38 & 0.12  &    0.24 &  0.30  &    0.582$^a$   & I$_{814}$ \\
     427 &  2718.8 & 2204.3 & 20.86  &   1.30 & 0.14  &    0.63 &  0.69  &     0.583$^a$    & I$_{814}$ \\
     645 &  1839.4 & 4121.8 & 20.55  &   0.38 & 0.19  &    1.00 &  1.50  &   0.600   & I$_{814}$ \\
      30 &  1189.3 &  478.1 & 21.45  &   0.37 & 0.25  &    0.26 &  1.40  &   0.620   & I$_{814}$ \\
     684 &  2099.1 & 3636.9 & 20.02  &   0.36 & 0.29  &    1.27 &  3.52  &   0.673$^a$   & I$_{814}$ \\
     297 &  1827.8 & 1697.9 & 22.06  &   0.21 & 0.30  &    0.17 &  0.75  &    0.680   & I$_{814}$ \\
     198 &  832.7  & 1263.0 & 21.61  &   0.43 & 0.32  &    0.57 &  0.30  &     0.696$^a$    & I$_{814}$ \\
      65 &   995.6 &  654.3 & 23.05  &   0.24 & 0.66  &    0.17 &  0.10  &    0.720   & I$_{814}$ \\
     394 &  3953.7 & 2367.7 & 20.39  &   0.35 & 0.37  &    2.51 &  0.09  &     0.760   & I$_{814}$ \\
     610 &  1674.7 & 3067.4 & 19.95  &   0.84 & 0.28  &    1.66 &  3.46  &   0.760$^a$   & I$_{814}$ \\
     768 &  2591.1 & 3492.9 & 23.36  &   0.30 & 0.67  &    0.13 &  0.08  &     0.760   & I$_{814}$ \\
     774 &  2782.2 & 3544.1 & 22.43  &   0.39 & 0.43  &    0.33 &  0.19  &    0.800   & I$_{814}$ \\
     777 &  1520.1 & 3463.8 & 22.06  &   0.34 & 0.24  &    0.60 &  0.25  &     0.800   & I$_{814}$ \\
     306 &  1129.4 & 1783.3 & 22.85  &   0.43 & 0.70  &    0.42 &  0.20  &    0.940   & J$_s$ \\
     337 &  3929.8 & 2643.0 & 21.82  &   0.39 & 0.08  &    1.08 &  0.74  &   0.940   & J$_s$ \\
     638 &  2661.3 & 4089.3 & 23.02  &   0.28 & 0.42  &    0.24 &  0.12  &    0.940   & J$_s$  \\
     795 &  1214.5 & 3434.2 & 23.37  &   0.24 & 0.32  &    0.21 &  0.10  &    0.940   & J$_s$  \\
      64 &  2166.0 &  660.9 & 22.25  &   0.28 & 0.10  &    0.54 &  0.48  &    0.960   & J$_s$  \\
     268 &  1288.8 & 1574.3 & 22.10  &   0.09 & 0.15  &    0.37 &  1.83  &   0.960   & J$_s$  \\
     367 &  3558.3 & 2506.5 & 23.23  &   0.60 & 0.16  &    0.27 &  0.18  &     0.960   & J$_s$  \\
     429 &   581.0 & 2211.8 & 21.79  &   0.75 & 0.73  &    0.82 &  0.68  &    0.970$^a$   & J$_s$  \\
      97 &  3335.8 &  784.7 & 20.85  &   0.10 & 0.02  &    2.08 &  5.25  &    0.980   & J$_s$  \\
     301 &  1390.5 & 1700.6 & 23.15  &   0.22 & 0.26  &    0.24 &  0.14  &    0.980   & J$_s$  \\
     326 &  1654.8 & 2678.6 & 22.16  &   0.19 & 0.11  &    0.40 &  1.77  &   0.980   & J$_s$  \\
     420 &  2836.4 & 2263.4 & 21.69  &   0.15 & 0.14  &    0.61 &  3.02  &    0.980   & J$_s$ \\
     691 &  3374.1 & 3822.9 & 22.63  &   0.24 & 0.23  &    0.39 &  0.39  &     0.980   & J$_s$ \\
     608 &  3122.9 & 4270.2 & 22.18  &   0.41 & 0.22  &    0.61 &  0.42  &     1.000   & J$_s$ \\
     665 &  2535.8 & 3888.0 & 21.64  &   0.44 & 0.42  &    0.70 &  3.58  &   1.000   & J$_s$ \\
     224 &  2397.5 & 1376.9 & 21.82  &   0.24 & 0.24  &    0.56 &  1.08  &   1.020   & J$_s$ \\
     753 &  2850.8 & 3602.0 & 22.89  &   0.22 & 0.17  &    0.41 &  0.29  &    1.020   & J$_s$ \\
   10008 &   342.7 & 1274.3 & 22.32  &   0.19 & 0.38  &    0.32 &  1.05  &   1.040   & J$_s$ \\
     152 &  1123.2 &  994.5 & 23.00  &   0.40 & 0.35  &    0.43 &  0.31  &    1.060   & J$_s$ \\
     241 &  2508.8 & 1585.0 & 21.71  &   0.46 & 0.36  &    0.74 &  1.56  &   1.060   & J$_s$ \\
      79 &  3159.3 &  746.3 & 21.48  &   0.46 & 0.44  &    1.24 &  1.38  &   1.080   & J$_s$ \\
      18 &  1361.7 &  404.4 & 21.20  &   0.30 & 0.11  &    1.09 &  3.28  &   1.100   & J$_s$ \\
     249 &  2222.5 & 1495.3 & 22.59  &   0.55 & 0.28  &    0.29 &  0.63  &	1.100	& J$_s$ \\
     565 &  2534.7 & 2935.8 & 20.75  &   0.47 & 0.32  &    2.31 &  2.92  &    1.114$^a$   & J$_s$ \\
     686 &  2474.7 & 3767.0 & 21.06  &   0.33 & 0.19  &    1.57 &  2.82  &   1.116$^a$   & J$_s$ \\
     493 &   271.4 & 1901.0 & 20.97  &   0.74 & 0.40  &    1.60 &  2.02  &   1.120   & J$_s$ \\
      45 &  3166.5 &  573.0 & 20.89  &   0.28 & 0.01  &    2.04 &  4.08  &   1.140   & J$_s$ \\
     206 &  3136.3 & 1280.3 & 22.71  &   0.36 & 0.25  &    0.67 &  0.33  &    1.152$^a$   & J$_s$ \\
     276 &  2216.7 & 1691.8 & 20.89  &   0.27 & 0.20  &    2.00 &  6.13  &    1.160   & J$_s$ \\
     644 &  3261.3 & 3967.9 & 22.67  &   0.21 & 0.00  &    0.40 &  2.24  &   1.160   & J$_s$ \\
     669 &  2412.0 & 3968.5 & 23.27  &   0.47 & 0.06  &    0.46 &  0.23  &    1.200   & J$_s$ \\
     404 &  3693.2 & 2330.4 & 22.75  &   0.41 & 0.26  &    0.65 &  0.59  &    1.220   & J$_s$ \\
      27 &   956.2 &  585.7 & 20.22  &   1.07 & 0.08  &    4.25 &  8.05  &    1.230$^a$   & J$_s$ \\
     251 &   765.4 & 1492.1 & 22.79  &   0.30 & 0.77  &    0.54 &  0.70  &    1.240   & J$_s$ \\
     254 &  2579.9 & 1701.3 & 20.31  &   0.27 & 0.13  &    4.96 &  7.81  &    1.270$^a$   & J$_s$ \\
     101 &  1314.3 &  781.9 & 22.23  &   0.35 & 0.35  &    1.21 &  1.43  &   1.280   & J$_s$ \\
     149 &  2849.3 &  951.4 & 23.18  &   0.23 & 0.16  &    0.29 &  0.74  &    1.280   & J$_s$ \\
     470 &  2606.8 & 1988.5 & 20.39  &   0.52 & 0.03  &    3.93 &  6.13  &   1.284$^a$   & J$_s$ \\
     502 &  1804.0 & 1838.1 & 23.20  &   0.46 & 0.66  &    0.33 &  0.47  &   1.300   & J$_s$ \\
     771 &  3494.8 & 3469.8 & 22.85  &   0.58 & 0.43  &    0.45 &  0.69  &     1.300   & J$_s$ \\
     145 &  1432.7 &  983.2 & 22.34  &   0.71 & 0.32  &    0.74 &  1.00  &   1.320   & J$_s$ \\
     395 &  2926.7 & 2377.6 & 22.65  &   0.24 & 0.13  &    0.90 &  0.85  &     1.320   & J$_s$ \\
     637 &  2746.6 & 4100.2 & 21.94  &   0.27 & 0.48  &    1.68 &  1.81  &   1.320   & J$_s$ \\
     199 &  1544.7 & 1257.0 & 21.68  &   0.31 & 0.18  &    1.29 &  6.27  &   1.340   & J$_s$ \\
     791 &  2767.9 & 3454.1 & 22.97  &   0.33 & 0.47  &    0.58 &  0.58  &    1.360   & J$_s$ \\
     437 &  3508.0 & 2173.2 & 23.15  &   0.51 & 0.52  &    0.58 &  0.58  &     1.380   & J$_s$ \\
     201 &  3955.7 & 1215.0 & 22.96  &   0.32 & 0.18  &    0.99 &  0.64  &    1.400   & J$_s$ \\
     408 &  1297.5 & 2325.2 & 23.08  &   0.17 & 0.20  &    0.75 &  0.56  &     1.400   & J$_s$ \\
     785 &  3438.1 & 3196.3 & 21.57  &   0.43 & 0.23  &    2.16 &  3.33  &   1.400   & J$_s$ \\
     751 &  2004.1 & 3687.6 & 23.13  &   0.21 & 0.13  &    0.69 &  0.80  &     1.420   & J$_s$ \\
     302 &   770.9 & 1799.8 & 21.55  &   0.94 & 0.10  &    2.94 &  3.36  &    1.439$^a$   & J$_s$ \\
      61 &  3804.3 &  592.7 & 23.02  &   0.34 & 0.25  &    0.58 &  0.68  &     1.440   & J$_s$ \\
     783 &  2674.2 & 3530.1 & 22.50  &   0.26 & 0.19  &    0.86 &  1.21  &   1.440   & J$_s$ \\
   10001 &  2954.3 &  782.1 & 21.53  &   0.59 & 0.24  &    2.50 &  3.16  &  1.440   & J$_s$ \\
     781 &  1304.1 & 3494.1 & 22.72  &   0.53 & 0.53  &    1.08 &  0.75  &    1.480   & J$_s$ \\
     620 &  1876.8 & 3623.2 & 22.16  &   0.29 & 0.12  &    2.27 &  1.48  &   1.558$^a$   & H \\
     628 &  3265.9 & 4251.8 & 22.37  &   0.13 & 0.10  &    1.15 &  3.18  &    1.580   & H \\
     675 &  1836.0 & 3395.8 & 22.22  &   0.32 & 0.18  &    1.61 &  2.17  &   1.600   & H \\
     724 &  2272.9 & 3717.9 & 23.34  &   0.17 & 0.37  &    0.56 &  0.75  &     1.620   & H \\
     583 &  3745.2 & 3064.7 & 22.90  &   0.12 & 0.19  &    0.88 &  4.36  &   1.640   & H \\
     349 &  1064.0 & 2559.4 & 23.17  &   0.40 & 0.28  &    1.06 &  0.51  &     1.680   & H \\
     233 &  1148.6 & 1369.4 & 23.38  &   0.07 & 0.44  &    1.03 &  0.70  &    1.720   & H \\
     754 &  2616.4 & 3525.4 & 23.16  &   0.26 & 0.36  &    0.73 &  1.80  &   1.760   & H \\
     267 &   966.0 & 1628.1 & 21.83  &   0.58 & 0.26  &    3.42 &  3.39  &   1.820   & H \\
     810 &  3062.8 & 3348.3 & 22.80  &   0.14 & 0.12  &    1.17 &  2.69  &   1.920   & H \\
     600 &  1843.5 & 3658.4 & 22.32  &   0.52 & 0.33  &    2.94 &  3.49  &   1.960   & H \\
     500 &  2364.1 & 1852.0 & 23.24  &   0.15 & 0.54  &    0.89 &  1.91  &   2.020   & H \\
     290 &   696.4 & 1684.2 & 21.95  &   0.27 & 0.11  &    4.66 &  2.54  &   2.025$^a$   & H \\
     257 &  3011.6 & 1696.2 & 22.10  &   0.61 & 0.43  &    3.75 &  2.27  &  2.027$^a$	& H \\
      21 &  2416.4 &  406.0 & 23.49  &   0.20 & 0.54  &    1.15 &  0.39  &   2.040   & H \\
      96 &   375.0 &  716.1 & 23.35  &   0.30 & 0.57  &    1.40 &  0.56  &    2.060   & H \\
     776 &  3562.4 & 3421.4 & 22.44  &   0.21 & 0.16  &    2.82 &  1.96  &   2.077$^a$   & H \\
     173 &  572.9  & 1026.0 & 23.23  &   0.35 & 0.23  &    1.41 &  0.86  &     2.140   & H \\
     496 &   373.2 & 1877.0 & 22.40  &   0.21 & 0.13  &    2.40 &  4.49  &   2.140   & H \\
     729 &  3327.2 & 3768.1 & 22.73  &   0.38 & 0.31  &    2.51 &  0.91  &     2.140   & H \\
     143 &   779.7 &  969.2 & 23.37  &   0.48 & 0.34  &    1.41 &  0.94  &     2.160   & H \\
     242 &  1139.5 & 1425.8 & 23.43  &   0.25 & 0.17  &    1.25 &  0.55  &    2.160   & H \\
     219 &  3888.0 & 1291.8 & 23.35  &   0.35 & 0.62  &    1.27 &  1.35  &   2.200   & H \\
     375 &  1210.7 & 2405.4 & 22.79  &   0.47 & 0.21  &    2.01 &  3.14  &     2.240   & H \\
     767 &  3297.6 & 3622.1 & 22.54  &   0.17 & 0.41  &    3.06 &  10.1  &   2.300   & H \\
     161 &  2577.8 & 985.6  & 23.42  &   0.10 & 0.40  &    1.30 &  5.55  &     2.340   & H \\
     591 &  3936.8 & 3251.8 & 19.56  &   0.18 & 0.16  &    1.49 &  0.94  &   2.400   & H \\
     176 &   722.5 & 1089.1 & 22.92  &   0.46 & 0.24  &    2.79 &  4.11  &    2.500   & H \\
     363 &   537.5 & 2521.3 & 22.41  &   0.78 & 0.24  &    4.72 &  2.00  &   2.500   & H \\
   10006 &  3127.6 &  985.8 & 23.32  &   0.18 & 0.23  &    2.43 &  1.41  &   2.652$^a$   & K$_s$ \\
     656 &  2535.7 & 4060.5 & 22.69  &   0.68 & 0.21  &    4.21 &  15.2  &  2.740   & K$_s$ \\
     452 &   567.0 & 2083.6 & 22.84  &   0.31 & 0.57  &    4.13 &  3.06  &   2.760   & K$_s$ \\
     806 &  3266.3 & 3256.9 & 22.67  &   0.26 & 0.30  &    4.91 &  1.76  &   2.789$^a$   & K$_s$ \\
     807 &  3305.0 & 3263.3 & 22.70  &   0.45 & 0.14  &    4.86 &  1.86  &   2.790$^a$   & K$_s$ \\
     657 &  3370.9 & 4068.2 & 22.53  &   0.57 & 0.26  &    5.94 &  3.51  &   2.793$^a$   & K$_s$ \\
     294 &   382.4 & 1662.5 & 23.34  &   0.49 & 0.10  &    2.81 &  1.95  &     2.820   & K$_s$ \\
     453 &  3945.9 & 2085.7 & 23.27  &   0.24 & 0.12  &    2.99 &  8.15  &    2.900   & K$_s$ \\
     534 &  3452.7 & 2959.4 & 22.78  &   0.28 & 0.31  &    5.35 &  4.52  &    3.000   & K$_s$ \\
     494 &  2562.2 & 1910.4 & 22.99  &   0.93 & 0.31  &    4.47 &  2.28  &     3.000   & K$_s$ \\
     465 &   360.7 & 2029.8 & 23.37  &   0.14 & 0.71  &    3.13 &  3.27  &   3.040   & K$_s$ \\
     397 &  2460.0 & 2379.0 & 23.41  &   0.32 & 0.46  &    3.17 &  5.91  &   3.080   & K$_s$ \\
     448 &  2508.9 & 2111.6 & 23.46  &   0.16 & 0.41  &    2.78 &  1.25  &   3.140   & K$_s$ \\
     622 &  3299.0 & 4292.6 & 23.07  &   0.24 & 0.37  &    3.90 &  1.95  &     3.140  & K$_s$ \\
      98 &  4055.0 &  718.8 & 23.09  &   0.16 & 0.16  &    4.65 &  9.66  &    3.160   & K$_s$ \\
     624 &  3738.7 & 4243.9 & 23.19  &   0.21 & 0.16  &    4.22 &  9.01  &  3.160$^a$	& K$_s$ \\
     813 &  2779.1 & 3315.0 & 23.31  &   0.29 & 0.43  &    3.14 &  1.39  &   3.240   & K$_s$ \\
      80 &  2926.6 &  703.1 & 22.71  &   0.40 & 0.11  &    8.88 &  3.71  &   3.840   & K$_s$ \\

\enddata

\tablecomments{Col. (1): Catalog identification numbers (see Labb\'e et al.
2003). Col. (2) and (3): X and Y pixel coordinate positions in the HDF--S
mosaic. Col (4): $K_s$--band total magnitudes. Col. (5): Circularized restframe
half--light radii (arcsec). The typical uncertainty on the size determination
is 35\%. Col (6): intrinsic (i.e. the recovered non--seeing affected)
ellipticity. Col (7): Restframe V--band luminosity. The typical uncertainty on
the luminosity determination is 30\%. Col (8): Stellar mass. Col (9): redshift (the
index $a$ mean spectroscopic $z$).  Col (10): Filter used to measure the size
of the galaxies}

\label{tabdata}
\end{deluxetable}

\clearpage
\begin{deluxetable}{ccccc}

\tablecolumns{11}
\tablewidth{0pc}

\tablehead{ L$_V$(10$^{10}$ h$^{-2}$ L$_\odot$)  & $\bar{r}_e(L,0)$ &
$\sigma(L,0)$ & z' & $\bar{r}_e(L,z=2.5)$/$\bar{r}_e(L,0)$\\ (1) & (2) & (3) & (4)
& (5)}

\startdata
3--10     & 6.45 & 0.36  & 3.91 & 0.25($\pm$0.10) \\
1--3      & 3.68 & 0.33  & 2.58 & 0.35($\pm$0.10) \\
0.3--1    & 2.07 & 0.40  & 1.55 &  -- \\
\hline
M(10$^{10}$ h$^{-2}$ M$_\odot$)  & $\bar{r}_e(M,0)$ & $\sigma(M,0)$ & z'
&$\bar{r}_e(M,z=2.5)$/$\bar{r}_e(M,0)$ \\
\hline
3--10     & 2.46 & 0.44  & 3.17 & 0.60($\pm$0.15)\\
1--3      & 1.94 & 0.49  & 2.52 & 0.80($\pm$0.20)\\
0.3--1    & 1.70 & 0.52  & 1.70 & --\\

\enddata 

\tablecomments{Properties of the bin selection for analyzing the evolution of
sizes in the FIRES data. Col. (1): the luminosity (mass) range of the bin. Col.
(2): SDSS mean size (in h$^{-1}$ kpc) at z$\sim$0 of galaxies with a luminosity
equal to the mean luminosity of the FIRES galaxies in the luminosity (mass)
range of Col. (1). Col. (3): Same than in  Col. (2) but with the SDSS
dispersion. Col. (4): Largest observable redshift  (for galaxies with
K$_S<$23.5) for the mean luminosity (mass) of the analyzed bin. Col. (5):
Implied size evolution at z=2.5 according to our analysis (see text for
details).} 

\label{tabbin}

\end{deluxetable}

\end{document}